\documentclass[11pt]{article}

\usepackage{graphicx}
\usepackage{amsmath}
\usepackage{amsfonts}
\usepackage{epsfig}
\usepackage{color}

\newcommand{\be}{\begin{equation}}
\newcommand{\ee}{\end{equation}}
\newcommand{\bes}{\begin{equation*}}
\newcommand{\ees}{\end{equation*}}
\newcommand{\beqn}{\begin{eqnarray}}
\newcommand{\eeqn}{\end{eqnarray}}
\newcommand{\beqns}{\begin{eqnarray*}}
\newcommand{\eeqns}{\end{eqnarray*}}
\newcommand{\lkr}{\left(}
\newcommand{\lkv}{\left[}
\newcommand{\rkv}{\right]}
\newcommand{\rkr}{\right)}
\newcommand{\lfi}{\left\{}
\newcommand{\rfi}{\right\}}

\newcommand{\fr}[1]{(\ref{#1})}

\newcommand{\vart}{\vartheta}
\newcommand{\ro}{\varrho}
\newcommand{\ph}{\varphi}
\newcommand{\del}{\delta}

\newcommand{\al}{\alpha}
\newcommand{\af}{\alpha}
\newcommand{\eps}{\epsilon}

\newcommand{\ga}{\gamma}
\newcommand{\te}{\theta}
\newcommand{\om}{\omega}
\newcommand{\lam}{\lambda}
\newcommand{\Up}{\Upsilon}

\newcommand{\sig}{\sigma}

\newcommand{\Lam}{\Lambda}
\newcommand{\Om}{\Omega}

\newcommand{\EE}{\ensuremath{{\mathbb E}}}

\newcommand{\II}{\ensuremath{{\mathbb I}}}

\newcommand{\PP}{\ensuremath{{\mathbb P}}}

\newcommand{\RR}{\ensuremath{{\mathbb R}}}

\newcommand{\Span}{\mbox{Span}}

\newcommand{\Var}{\mbox{Var}}

\newcommand{\diag}{\mbox{diag}}

\newcommand{\Tr}{\mbox{Tr}}
\newcommand{\proj}{\mbox{proj}}

\newtheorem{theorem}{Theorem}
\newtheorem{lemma}{Lemma}
\newtheorem{corollary}{Corollary}

\newtheorem{remark}{Remark}

\newcommand{\ba}{\mathbf{a}}

\newcommand{\bd}{\mathbf{d}}
\newcommand{\bof}{\mathbf{f}}
\newcommand{\bh}{\mathbf{h}}
\newcommand{\bq}{\mathbf{q}}
\newcommand{\bt}{\mathbf{t}}
\newcommand{\bu}{\mathbf{u}}
\newcommand{\bv}{\mathbf{v}}

\newcommand{\by}{\mathbf{y}}

\newcommand{\bA}{\mathbf{A}}

\newcommand{\bD}{\mathbf{D}}
\newcommand{\bI}{\mathbf{I}}
\newcommand{\bM}{\mathbf{M}}
\newcommand{\bQ}{\mathbf{Q}}

\newcommand{\bU}{\mathbf{U}}

\newcommand{\bW}{\mathbf{W}}
\newcommand{\bX}{\mathbf{X}}
\newcommand{\bY}{\mathbf{Y}}

 \newcommand{\bte}{\mbox{\mathversion{bold}$\te$}}
\newcommand{\bxi}{\mbox{\mathversion{bold}$\xi$}}
\newcommand{\boeta}{\mbox{\mathversion{bold}$\eta$}}
\newcommand{\bbe}{\mbox{\mathversion{bold}$\beta$}}

\newcommand{\bph}{\mbox{\mathversion{bold}$\ph$}}

\newcommand{\bobeta}{\mbox{\mathversion{bold}$\beta$}}
\newcommand{\bgamma}{\mbox{\mathversion{bold}$\gamma$}}

\newcommand{\bPhi}{\mbox{\mathversion{bold}$\Phi$}}
\newcommand{\bUp}{\mbox{\mathversion{bold}$\Up$}}
\newcommand{\bPsi}{\mbox{\mathversion{bold}$\Psi$}}

\newcommand{\Jc}{{J^c}}

\newcommand{\calD}{{\mathcal{D}}}
\newcommand{\calG}{{\mathcal G}}
\newcommand{\calJ}{{\mathcal{J}}}

\newcommand{\calL}{{\mathcal{L}}}
\newcommand{\calP}{{\mathcal{P}}}
\newcommand{\calX}{{\cal{X}}}
\newcommand{\calY}{{\cal{Y}}}

\newcommand{\calH}{{\cal H}}

\newcommand{\calN}{{\cal N}}

\newcommand{\fhbte}{f_{\widehat{\bte}}}
\newcommand{\hbte}{\widehat{\bte}}
\newcommand{\hte}{\widehat{\te}}

\newcommand{\hbf}{\hat{\bof}}
\newcommand{\hbq}{\hat{\bq}}

\newcommand{\hk}{\widehat{k}}
\newcommand{\tk}{\widetilde{k}}

\newcommand{\lan}{\langle}
\newcommand{\ran}{\rangle}

\newcommand{\alfo}{\al_0}

\newcommand{\sumjp}{\sum_{j=1}^p}
\newcommand{\sumin}{\sum_{i=1}^n}
\newcommand{\sumJ}{\sum_{j \in J}}

\newcommand{\phj}{\ph_j}
\newcommand{\psij}{\psi_j}
\newcommand{\nuj}{\nu_j}
\newcommand{\tej}{\te_j}

\newcommand{\bej}{\beta_j}

\newcommand{\hbejdel}{\widehat{\beta}_{j,\del}}
\newcommand{\hbej}{\widehat{\beta}_{j}}

\newcommand{\phjdel}{\ph_{j, \del}}
\newcommand{\psijdel}{\psi_{j, \del}}
\newcommand{\nujdel}{\nu_{j, \del}}

\newcommand{\tpsi}{\widetilde{\psi}}
\newcommand{\tphi}{\widetilde{\ph}}

\newcommand{\Cups}{C_{\nu}}
\newcommand{\Dxi}{\Delta x_i}

\newcommand{\lamin}{\lam_{\min}}
\newcommand{\lamax}{\lam_{\max}}

\newcommand{\ft}{f_{\bt}}
\newcommand{\fte}{f_{{\bte}}}

\newcommand{\hf}{\widehat{f}}

\newcommand{\hbobeta}{\widehat{\bobeta}}
\newcommand{\deljo}{\del_{j1}}
\newcommand{\deljt}{\del_{j2}}

\begin{document}

\title{\bf { Solution of linear   ill-posed problems using overcomplete dictionaries }}

\author{{\em   Marianna Pensky}   \\
         Department of Mathematics,
         University of Central Florida 
\vspace{4mm}\\
  {\bf In memory of Laurent Cavalier  } }

\date{}

\bibliographystyle{plain}
\maketitle

\begin{abstract}
In the present paper we consider application of   overcomplete dictionaries to solution of general ill-posed linear inverse problems. 
 Construction of an adaptive optimal solution for problems of this sort usually relies either   on a singular value decomposition (SVD)
or  representation of the solution via some orthonormal basis. The shortcoming of both approaches   lies
in the fact that, in many situations,  neither the eigenbasis of the linear operator nor a standard orthonormal basis 
 constitutes an appropriate collection of functions for sparse representation of the unknown function.

In the context of regression problems, there have been an enormous amount of effort to recover an unknown function
using an  overcomplete dictionary. One of the most popular methods, Lasso and its versions, is based 
on minimizing the empirical likelihood and  requires stringent assumptions on the dictionary, the, so called, compatibility conditions.  
While these conditions may be satisfied for the original dictionary functions, they usually do not hold for their images   
due to contraction imposed by the linear operator.

In what follows, we bypass this difficulty by a novel approach which  is based on inverting   each of the dictionary functions  and matching 
the resulting expansion to the true function, thus, avoiding unrealistic assumptions on the dictionary and using Lasso in a predictive setting.
We examine both the white noise and the observational model formulations and also discuss how  exact 
inverse images of the dictionary functions can be replaced by their approximate counterparts. Furthermore, we  show how the suggested methodology 
can be extended to the problem of estimation of a mixing density in a continuous mixture. For all the situations listed above, 
we provide the oracle inequalities for the risk in a finite sample setting. 

We compare the performance of the suggested methodology with the estimators based on the SVD and the orthonormal basis decomposition 
as well as with the wavelet-vaguelette estimator. Simulation studies confirm good computational properties of the Lasso-based technique.

\vspace{2mm} 


\vspace{2mm}{\bf AMS (2000) Subject Classification}: {Primary: 62G05.  Secondary:  62C10  }
\end{abstract}

\section{Introduction  }
\label{sec:introduction}
\setcounter{equation}{0}

In this paper, we consider   solution of a general ill-posed linear inverse problem $Qf = q$ where $Q$ is a  
bounded linear  operator that  does not have a bounded inverse and the right-hand side $q$ is measured with error. 
Problems of this kind appear in many areas of application such as astronomy (blurred images), econometrics 
(instrumental variables), medical imaging (tomography, dynamic contrast enhanced CT and MRI), 
finance (model calibration of volatility) and many others.

In particular, we consider equation 
\be \label{geneq}
y  =  q  + \sqrt{\eps} \eta,\quad q  =  Q f,  
\ee
where $\eta(t)$ is the Gaussian process representing the noise,  $\sqrt{\eps}$ is the noise level  and 
$Q:  \calH_1   \to \calH_2$ is a bounded linear operator which does not have a bounded inverse, so problem \fr{geneq} is indeed ill-posed. 
Here,  $\calH_1$ and $\calH_2$ are Hilbert spaces. We assume that observations are  taken as 
 functionals of $y$
\be \label{eq:observ}
\lan y, g \ran_{\calH_2} = \lan Qf, g \ran_{\calH_2} + \sqrt{\eps}\,  \eta(g), \ \ g \in \calH_2,
\ee
where, for any $g \in \calH_2$,  $\eta(g)$ is a Gaussian random variable with zero  mean  and variance $\|g\|_2^2$ such that
$\EE [\eta(g_1) \eta(g_2)] = \lan g_1, g_2 \ran_{\calH_2}$. 
Formulation above refers to the scenario  where one cannot measure function $y(t)$ everywhere:   only 
functionals of $y$ are available. Such functionals, for example,  may be coefficients of $y$ in some orthonormal basis 
(e.g., Fourier, wavelet, eigenbasis of $Q$). The situation where one observes   values of function $y$ at some points  is 
studied in depth in Section \ref{sec:obser_model}.

In order to understand formulation above, consider a common situation where 
operator $Q$ is of the form
\be \label{Q_example}
(Qf) (x) = \int_a^b g(x,t) f(t) dt, \quad x \in  (c,d).
\ee 
and $f(t)$, $g(x,t)$,  $q(x)$ and $y(x)$, $t \in (a,b)$, $x \in (c,d)$, are 
square integrable  functions. 
In this case, $Q:  \calH_1   \to \calH_2$ where $\calH_1 = L^2(a,b)$ and 
$\calH_2 = L^2(c,d)$, the spaces of square integrable functions with the  scalar products  
\bes
\lan f_1, f_2 \ran_{\calH_1} = \int_a^b f_1 (t) f_2 (t) dt, \quad
\lan q_1, q_2 \ran_{\calH_1} = \int_c^d q_1 (x) q_2 (x) dx.
\ees
Formula \fr{eq:observ} refers to the fact that function $y(x)= (Qf)(x)   + \sqrt{\eps} \eta (x)$   cannot be measured for all points $x$:
one can only observe linear functionals
\bes
\lan y, \psi \ran_{\calH_2} = \int_c^d y(x) \psi(x) dx = \int_c^d q(x) \psi(x) dx 
+ \sqrt{\eps}\,  \eta(\psi),
\ees  
where  $\eta(\psi) \sim N \lkr 0, \|\psi\|_2^2 \rkr$.

Solutions of statistical inverse problem  \fr{geneq}   usually rely on reduction of the problem 
to the sequence model by   carrying out the singular value decomposition (SVD)  (see, e.g., \cite{cavalgol2}, \cite{cavalgol1},
\cite{cavreiss},  \cite{gol} and \cite{kalifa}), or its relaxed version, the 
wavelet-vaguelette decomposition proposed by Donoho \cite{donoho} and further studies by Abramovich and Silverman \cite{abr}. 
Another general approach  is Galerkin method with subsequent model selection  (see, e.g.,    
 \cite{cohen}, \cite{efrom} and \cite{hoffmann}).

The advantage of the methodologies listed above is that they are asymptotically optimal in a minimax 
sense and, hence, deliver the best possible rates in the ``worst case scenario'' setting.
The function of interest is usually represented via an   orthonormal basis which is
motivated by the form of the operator $Q$. However, in spite of being minimax optimal in many contexts, 
these approaches   have two drawbacks. The first one  is that, in many situations, these techniques may 
not be applicable. Indeed,  for majority of linear operators, the SVD decomposition 
is unknown and, hence, cannot be applied. Wavelet-vaguelette decomposition relies on 
relatively stringent conditions that are satisfied only for 
specific operators, mainly, of convolution type. In addition, wavelet-based methods
are advantageous when one recovers a one-dimensional function defined on a finite interval but do not perform as well 
for a function  of several variables or with an infinite domain. 
Another shortcoming is that the orthonormal dictionary may not be "rich enough". 
If the unknown function does not have a relatively compact and accurate representation 
in the chosen basis,  the accuracy of the resulting estimator will be  poor even though
the inverse image of $f$ has a moderate norm. In Section \ref{sec:advantages} we provide 
detailed explanations how application of overcomplete dictionaries can improve 
precision of the solutions of  ill-posed linear inverse problems.

In the last decade, a great deal of effort was spent on recovery of an unknown function $f$ in regression setting from its noisy observations 
using overcomplete  dictionaries. In particular, if the dictionary is large enough and $f$ has a sparse representation in this dictionary,
then $f$ can be recovered with a much better precision than, for example,  when  it is expanded over an orthonormal basis. 
The methodology is based on the idea  that the error of an estimator of $f$  is approximately
proportional  to the number of dictionary functions that are used for   representing   $f$,  
therefore,  expanding a function of interest over fewer dictionary elements decreases 
the estimation error.  In order to represent a variety of functions efficiently, one would need to consider a dictionary of much larger 
size than the number of available  observations, the, so called, overcomplete dictionary, and also to develop  tools for 
choosing correct  elements of the dictionary that deliver efficient representation of $f$.

A variety of techniques have been developed for solution of those problems such as likelihood  
penalization methods and greedy algorithms. The most popular of those methods (due to its computational convenience),  
Lasso and its versions,  have been used  for solution of a number of theoretical and applied statistical 
problems (see, e.g.,  \cite{bickel_ritov_tsybakov},  \cite{bunea_tsybakov_2},  \cite{arnak},
 \cite{lounici_pontil_tsybakov}, \cite{yuan}, and also \cite{sara} and references therein).
However, application of Lasso is based on minimizing the empirical likelihood and, unfortunately, requires 
stringent assumptions on the dictionary $\lfi \ph_k \rfi_{j=1}^p$, the, so called, compatibility conditions. 
In regression set up, as long as compatibility conditions hold, one can identify 
the dictionary elements which represent the function of interest 
best of all at a ''price'' which is proportional to $\sqrt{\eps\, \log p}$  where $p$ is the dictionary size.
Regrettably, while compatibility conditions may be satisfied for the functions $\ph_j$ in the  original dictionary,
they usually do not hold for their images $Q \ph_j$  due to contraction imposed by the   operator $Q$.

In order to illustrate this issue, expand  $f$ over  the dictionary as $\fte  = \sum_{j=1}^p  \te_j  \ph_j $.
Then, $q_{\bte} = \sum_{j=1}^p  \te_j  u_j $ with $u_j = Q \ph_j$. In a nutshell,   in order Lasso  can recover  
vector of coefficients $\bte$ correctly,  matrix $\bA$ with elements $A_{kj} = \lan u_k, u_j \ran$
  should be such that its  sub-matrices of a small order  have eigenvalues that are uniformly separated 
from zero and infinity (see, e.g. \cite{bickel_ritov_tsybakov}). The latter   usually does not hold for  the
ill-posed problems where the smallest eigenvalue can decrease polynomially or even exponentially as a function of $j$.

The objective of this paper is to  circumvent this difficulty and apply Lasso methodology to 
solution of linear inverse problem \fr{geneq}.  For thus purpose, in Section \ref{sec:method},  instead of matching 
the expansion $q_{\bte}$ to data $y$, we invert each of the dictionary functions $\ph_j$ 
and match expansion $\fte$ to the true function $f$. This approach has several advantages. First, it allows 
to use  Lasso in a prediction setting where it requires much milder assumptions. In this setting,  
Lasso  converges to the true solution, although at a   slow rate,   under practically no assumptions on the dictionary. 
Second, inverting fully known functions $\ph_j$ is an easier task than inverting an unknown function  measured with noise.
In addition, the norms of the inverted images can be viewed as  a ``price'' of including each of the dictionary functions~$\ph_j$.
In order to ensure that the estimator $\fhbte$ attains fast convergence rates, we formulate a compatibility assumption and discuss 
sufficient conditions that guarantee its validity.

The Lasso methodology developed for equations \fr{geneq} and \fr{eq:observ} 
allows   a variety of generalizations.  First, in Section~\ref{sec:obser_model},
we extend formulations \fr{geneq} and \fr{eq:observ}  to observational model where 
only the values$y(t_i)$, $i=1, \cdots, n$, of $y(t)$  are available. Second, in Section~\ref{sec:mix_den}, we explain 
how, with very minor modifications, the Lasso technique can be used for estimation of a mixing density 
in a continuous mixture. Third, in Section \ref{sec:approx_images},  we show that,   
even if the exact inverse images of the dictionary functions do not exist, one can use 
their approximations and take advantage of the exact knowledge of the  dictionary functions 
which allows the optimal bias-variance decomposition.

We would like to emphasize  that the Lasso methodology for solution of linear inverse problems can be viewed as an extension 
of both the Galerkin method and the wavelet-vaguelette decomposition. Really, if instead of an overcomplete dictionary,
one uses an orthonormal basis, then Lasso methodology just reduces to Galerkin method with model selection carried out by 
a soft thresholding technique. Moreover, if this orthonormal basis is comprised of wavelet functions and 
conditions for validity of the wavelet-vaguelette decomposition hold, Lasso penalty just imposes soft thresholding
on the  wavelet coefficients.  In order to compare the Lasso estimator with those techniques, 
we carried out a  numerical study of the  the Laplace deconvolution problem considered, as an example,  
in Section~\ref{sec:Laplace}. In particular, together with the Lasso estimator, we  implemented   the SVD, 
the wavelet-vaguelette and the Laguerre basis   based estimators. 
 Simulation studies confirm  that the Lasso estimator developed in the paper has good precision.

The rest of the paper is organized as follows.
After introducing notations  (Section~\ref{sec:notations}),   Section \ref{sec:advantages}  explains
why application of overcomplete dictionaries allows to improve estimation precision in linear   ill-posed problems.
Section~\ref{sec:method} develops the theoretical foundations of the paper  by justifying application of   
  Lasso technique to solution of    general linear inverse problem \fr{geneq}.  In particular, it 
 introduces  a compatibility assumption which guarantees that the Lasso estimator   attains fast convergence rates
for any function $f$ which has a sparse representation in the dictionary. 
Section~\ref{sec:compat},   discusses this compatibility assumption and formulates simpler sufficient conditions
under which it holds. Sections~\ref{sec:obser_model}~and~\ref{sec:mix_den} clarify how this theory   
can be applied to the real life observational model and also to  estimation of  a mixing density on the 
basis of observations of a continuous mixture. Section~\ref{sec:approx_images} demonstrates how   exact 
inverse images of the dictionary functions can be replaced by their approximate counterparts.
Section~\ref{sec:examples} contains examples of applications of Lasso to the models studied in the previous sections. 
Section~\ref{sec:simulation} presents  a   simulation study.
Section~\ref{sec:discussion} concludes the paper with discussion of the results. 
Finally, Section~\ref{sec:proofs} contains proofs of the statements formulated in earlier  sections.

\section{Notations} 
\label{sec:notations}
\setcounter{equation}{0}

In the paper, we use the following notations.

\begin{itemize}

\item
For any vector $\bt \in \RR^p$, denote  its $\ell_2$, $\ell_1$, $\ell_0$ and $\ell_\infty$ norms by, 
respectively,  $\| \bt\|_2$, $\| \bt\|_1$,  $\| \bt\|_0$ and $\| \bt\|_\infty$.
Similarly, for  any function $f$,   denote by $\| f \|_2$, $\| f\|_1$ and 
$\| f \|_{\infty}$ its $L_2$, $L_1$ and $L_{\infty}$ norms.

\item 
For any matrix $\bA$,  denote its spectral and Frobenius norms by, respectively,  $\| \bA \|$ and $\| \bA \|_2$.
Notation $\bA >0$ or $\bA \geq 0$ means, respectively,  that $\bA$ is positive or non-negative definite.
Denote determinant of $\bA$ by $|\bA|$ and the largest, in absolute value, element of $\bA$ by $\| \bA\|_{\infty}$.
Denote the Moore-Penrose inverse of matrix $\bA$ by $\bA^{+}$.

\item
Denote $\calP = \{1, \cdots, p\}$.   For any subset of indices  $J \subseteq \calP$, 
 subset $J^c$ is its complement in $\calP$ and  $|J|$ is its cardinality, so that $|\calP| =p$. 
Let  $ \calL_J = \Span \lfi \ph_j, \   j \in J \rfi$. 

\item
If $J \subset \calP$ and $\bt \in \RR^p$,  then     $\bt_J \in \RR^{|J|}$ denotes   reduction of vector $\bt$ to
subset of indices $J$.

\item
Denote by $\lamin (m; \bPhi)$ and $\lamax (m; \bPhi)$ the minimum and the maximum restricted eigenvalues
of matrix $\bPhi$ 
\be \label{eigrestrict}
\lamin (m; \bPhi) = \min_{\stackrel{\bt \in \RR^p}{\|\bt \|_0 \leq m}}\ \frac{\bt^T \bPhi \bt}{\| \bt \|_2^2}, 
\quad
\lamax (m; \bPhi) = \max_{\stackrel{\bt \in \RR^p}{\|\bt \|_0 \leq m}}\ \frac{\bt^T \bPhi \bt}{\| \bt \|_2^2}. 
\ee
Also,   denote by $\varrho (\bPhi)$ the maximum of a non-diagonal  element of matrix $\bPhi$:
\be \label{nondiag}
 \varrho (\bPhi) = \max_{j \neq k} |\Phi_{jk}|.  
\ee
Whenever there is no ambiguity, we drop $\bPhi$ in the above notations and write simply  
$\lamin (m)$, $\lamax (m)$ and $\varrho$.

\item
 $a_m \asymp b_m$ means that there exist    constants $0 < C_1<C_2<\infty$
independent of $m$ such that $C_1 a_m < b_m < C_2 a_m$.
\end{itemize}


\section{Advantages of overcomplete dictionaries}
\label{sec:advantages}
\setcounter{equation}{0}

The purpose of this section is to demonstrate how application of a rich overcomplete dictionary can reduce 
estimation error in inverse  linear  ill-posed problems. 
Indeed, if an overcomplete dictionary allows an efficient representation of $f$, 
it leads to a smaller estimation error. In order to understand the roots of this phenomenon,
consider the situation where operator $Q$ has a singular value decomposition 
$Q e_k = \lam_k e_k$, $k=1, 2, \cdots$, and function $f$ can be represented as 
$f = \sum_k c_k e_k$. Assume, without loss of generality, that for some $\mu >0$ and $\rho >0$
\be \label{cklamk}
|c_k| \leq C_0 k^{-(\mu+1/2)}\quad \mbox{and} \quad |\lam_k| \geq C_\lam k^{-\rho}.
\ee
In this case, one can construct the SVD estimator $\hf_{SVD} = \sum_{k=1}^m \lam_k^{-1} \lan y, e_k \ran\, e_k$
 of $f$ with the mean squared error (MSE) of the form
\be \label{svd_er}
 \EE \|\hf_{SVD} - f \|_2^2 = \sum_{k=m+1}^{\infty}  c_k^2 + \eps  \sum_{k=1}^m \lam_k^{-2} 
 \asymp m^{-2\mu} + \eps  m^{1 + 2 \rho} \asymp \eps^{\frac{2\mu}{2\mu +2 \rho +1}},
\ee
where  the   value of $m$ is chosen to minimize the right-hand side of \fr{svd_er}.  
The advantage of the SVD is that its error rates hold in the ``worst case'' minimax estimation scenario  
where $f$ is the hardest to estimate in the chosen class of functions.

On the other hand, consider the ``best case'' scenario when one has an extensive overcomplete dictionary 
$\ph_l$ with $\| \ph_l\|=1$, $l=1, \cdots, p$,  and $f$ is proportional to one of the dictionary functions, say, $\ph_j$.
Expand  dictionary functions $\ph_l$ in the eigenbasis $e_k$ and find their inverse images $\psi_l$ obtaining
\bes 
\ph_l= \sum_{k=1}^\infty c_{lk} e_k, \quad
\psi_l = \sum_{k=1}^\infty c_{lk} \lam_k^{-1} e_k.
\ees
If one had an oracle which identifies  the function  $\ph_j$ that is proportional to $f$, 
then $c_{jk} = c_k/\|f\|$ and $f$ would be estimated by 
$\hf_{or} = \langle y, \psi_j \rangle \ph_j$ with the error  
 \beqn \label{eq:errorrate}
\EE \|\hf_{or} - f \|_2^2 & = &   \eps \| \psi_j \|^2_2 = \eps\, \sum_{k=1}^\infty \lam_k^{-2} c_{jk}^2 
= \eps\,  \|f\|_2^{-2}\, \sum_{k=1}^\infty \lam_k^{-2} c_k^2. 
\eeqn 
Moreover, if $\mu >  \rho$ in \fr{cklamk}, then the series in the right-hand side of \fr{eq:errorrate}
is convergent and $\hf_{or}$ has parametric error  rate $\EE \|\hf_{or} - f \|^2 \asymp \eps.$
Otherwise, if $\mu \leq  \rho$, one can replace $\psi_j$ by  
\be \label{psijmj}
\psi_{j, M_j} = \sum_{k=1}^{M_j} c_{jk} e_k
\ee 
and estimate $f$ by  $\hf_{or, M} = \langle y, \psi_{j, M_j} \rangle \ph_{j}$.
It is easy to calculate that 
\beqn \label{hforM}
\EE \|\hf_{or, M_j} - f \|_2^2 & = &    \lkv \eps\,  \sum_{k=1}^{M_j} \lam_k^{-2} c_{jk}^2 + \sum_{k = M_j+1}^\infty c_{jk}^2 \rkv 
\asymp \eps {M_j}^{2\rho - 2 \mu} + {M_j}^{-2 \mu}.
\eeqn 
Choosing $M_j$ that minimizes the right-hand side of \fr{hforM},   obtain 
\bes
\EE \|\hf_{or,M_j} - f \|_2^2 \asymp \eps^{\frac{2\mu}{2 \rho}} = o \lkr \eps^{\frac{2\mu}{2\mu +2 \rho +1}} \rkr 
\asymp \EE \|\hf_{SVD} - f \|_2^2,\quad \eps \to 0,
\ees
i.e. the error of $\hf_{or,M_j}$ is smaller than the error of the SVD estimator. 
The advantage comes from the fact that, unlike in \fr{svd_er}, in the right-hand sides of \fr{eq:errorrate}
and \fr{hforM}, the ``large'' values $\lam_k^{-2}$ 
are multiplied by ``small'' values $c_{jk}^2$ in the expression for the MSE.

One would argue that the assumption that $f$ is proportional to one of the dictionary elements is not very 
realistic. However, it is very likely that $f$ can be represented by a small subset of the dictionary functions 
$\ph_j, j \in J,$ of cardinality $|J| = s$. Then, $f$ can be estimated by 
\bes 
\hf_{or, \bM} = \sum_{j \in J} \langle y, \psi_{j, M_j} \rangle \ph_j,
\quad \bM = (M_1, \cdots, M_s)
\ees
where $\psi_{j, M_j}$ are defined in \fr{psijmj} and the values $M_j$ are found by 
minimizing the right-hand side of \fr{hforM}. If, for example,  the dictionary  functions are 
not ``much harder'' than $f$, i.e., if there exists a constant $C_f$ such that for $j \in J$ one has $c_{jk}^2 \leq C_f c_k^2$, 
then  $\EE \|\hf_{or,\bM} - f \|^2 \asymp s \eps$ if $\mu >  \rho$ and 
$\EE \|\hf_{or,\bM} - f \|^2 \asymp s \eps^{\frac{2\mu}{2 \rho}}$ otherwise.
Note that there is  also a significant difference between choosing the optimal values of $m$ in \fr{svd_er}
and $M_j$ in \fr{hforM}. Indeed, the coefficients of the dictionary functions $c_{jk}$ are known,
while coefficients $c_k$ of $f$ are unknown, so the latter problem is a straightforward one while the former one is not.

Since one does not have an oracle which allows to choose the ``right'' subset of dictionary functions $\ph_j, j \in J$,
Lasso is instrumental for choosing an appropriate subset such that, even if it does not coincide with the ``true'' subset $J$,
it provides an estimator of a similar quality.


\section{Lasso solution of a general linear inverse problem}
\label{sec:method}
\setcounter{equation}{0}

Consider equation \fr{geneq} described above with observations defined in \fr{eq:observ}.
Denote by $Q^*$ the conjugate operator for $Q$,
so that $\lan Qf, g \ran_{\calH_2}  = \lan f, Q^* g \ran_{\calH_1}$  for any $f \in \calH_1$ 
and $g \in \calH_2$. Unless there is an ambiguity, in what follows, 
we   denote the scalar product induced norms in both $\calH_1$ and $\calH_2$ by $\| \cdot \|_2$.

Let $\lfi \ph_j, j \in \calP \rfi$ be a dictionary  such that $\| \ph_j \|_{\calH_1} =1$.
%
Denote by $f$ the true solution of the problem \fr{geneq}
and by $f_{\bte}$   the   projection of this true 
solution on the linear span of functions $\lfi \ph_j, j \in \calP \rfi$ where, for any $\bt \in \RR^p$, we denote
 \be \label{f_expan}
\ft   = \sum_{j=1}^p  t_j  \ph_j.
\ee
If function $f$ were known, we would search for the vector of coefficients  $\bte$ of $f_{\bte}$  as a solution of the  
 optimization problem
  \bes 
 \bte = \arg\min_{\bt}    \| f - \ft \|_2^2,    
  \ees
where $\ft$ is defined in \fr{f_expan}. Note that, although $f$ is unknown,  
 \be \label{ernorm}
\| f - \ft \|_2^2 = \| f \|_2^2 + \| \ft \|_2^2 - 2 \sumjp \lan  f, \ph_j  \ran_{\calH_1}  t_j  
 \ee
is the sum of   three components where the first one,  $\| f \|_2^2$, is independent of $\bt$, 
and the second one, $\| \ft \|_2^2$, is completely known. In order to 
estimate the last term in \fr{ernorm},  we assume that the following condition holds:
\\

\noindent
{\bf (A0)} \ There exist $\psij \in \calH_2$ such that $Q^* \psij = \phj$ and 
$\nuj = \| \psi_j \|_{\calH_2}  < \infty$.
\\

\noindent
For example, if operator $Q$ is defined by formula \fr{Q_example}, then $\psij$ in Assumption~{\bf (A0)} 
are solutions of the following equations
\be \label{Q*_example}
(Q^* \psij) (t) = \int_c^d g(x,t) \psij(x) dx = \phj(t), \quad t \in  (a,b).
\ee 
Observe that equations resulting from Assumption~{\bf (A0)} have completely known right-hand sides.
The values of  $\nuj$ can be viewed as the ``price'' of estimating coefficient $\tej$ of $f_{\bte}$.
While, in the regression set up, this ``price'' is uniform for all coefficients, this is no longer true 
in the case of ill-posed problems. 
Under Assumption {\bf A0}, one can write 
\bes  
\bej = \lan  f,  \ph_j  \ran_{\calH_1} = \lan   f, Q^* \psij \ran_{\calH_1} = \lan   Qf,   
\psij \ran_{\calH_2} = \lan   q,   \psij \ran_{\calH_2},
\ees 
so that 
\be \label{mainrel}
\bej = \EE \lan   y,   \psij \ran_{\calH_2}. 
\ee
For this reason, we can replace $\bej = \lan  f,  \ph_j  \ran_{\calH_1}$  in \fr{ernorm} by its estimator 
\be \label{hbej1}
\hbej = \lan   y,   \psij \ran_{\calH_2}
\ee
and estimate the vector of coefficients $\bte$ by  
\be \label{las_sol}
\hbte = \arg\min_{\bt}    \lfi     \| \ft \|_{2}^2  - 2 \sumjp \hbej t_j + 
\af \sumjp \nu_j |t_j|   \rfi.
\ee
Note that \fr{las_sol} is the weighted Lasso problem  with the penalty  parameter $\af$.
The coefficients $\nuj$ in front of $|t_j|$ are motivated by the fact that $\hbej$ are centered normal 
variables with the variances $\nuj^2= \| \psij \|^2_2$.

In order to reduce optimization problem \fr{las_sol} to familiar matrix formulation,
we introduce matrix $\bPhi$ with elements $\Phi_{jk} = \lan \ph_j, \ph_k \ran$
and vector $\hbobeta$ with elements $\hbej$. Define matrices  $\bW$ and $\bUp$ by
\be \label{bWbUp}
\bW^T \bW = \bPhi, \quad \bUp = \diag(\nu_1, \cdots, \nu_p). 
\ee
 Then, \fr{las_sol} can be re-written as 
\be \label{las_sol1}
\hbte = \arg\min_{\bt}    \lfi   \bt^T \bW  \bW^T \bt -  2  \bt^T \hbobeta  + 
\af \| \bUp \bt \|_1   \rfi. 
\ee
Introducing vector $\bgamma$ such that $\bW^T \bgamma = \hbobeta$ we reduce \fr{las_sol1} to
\be \label{las_sol2}
\hbte = \arg\min_{\bt}    \lfi    \|\bW \bt -  \bgamma \|^2_2 + 
\af \| \bUp \bt \|_1   \rfi \quad \mbox{with} \quad   \bgamma =    (\bW \bW^T)^{+}  \bW \hbobeta.
\ee
Here, $\| \bUp \bt \|_1$ is the weighted Lasso penalty, $\alpha$ is the penalty parameter 
and $\bgamma$ is the right-hand side. 
 The  choices  of parameter $\alpha$ are discussed at the end of this section in Remark \ref{Las_param}.

Since we are interested in recovering $f$ rather that $\bte$ itself, 
we are using Lasso  for solution of the  prediction problem where it requires milder conditions on the dictionary. 
In particular, estimator $f_{\hbte}$ converges to the true function  $f$  with no additional assumptions 
on the dictionary.

\begin{theorem} \label{th:slow_Lasso}
Let Assumption {\bf  A0} hold. Then, for    any $\tau >0$  and any $\af \geq \alfo$, with probability at least
$1 - 2 p^{-\tau}$,  one has
\be \label{slowlas}
\| f_{\hbte} - f \|_2^2 \leq \inf_{\bt } \lkv  \| f_{\bt } - f \|_2^2 + 4 \al   \sumjp \nuj |t_j|  \rkv
\ee
where 
\be \label{alf0}
\alfo =   \sqrt{2\, \eps\, (\tau +1) \log p}.
\ee
\end{theorem}

\noindent
If the dictionary is large enough, so that $f_{\bte} = f$   where vector $\bte$ has support $J$ 
of size $|J| =s$ and   components of $\bte$  are uniformly bounded, 
then, with high probability, the error of estimating $f$ by $f_{\hbte}$ is  
$ \| f_{\hbte} - f \|_2^2  \asymp \sqrt{ \eps\,   \log p}\,  \sumJ   \nu_j $.
In the case of regression problem, $\nu_j =1$, so that    convergence rate appears as 
$ s  \sqrt{ \eps\,   \log p}$ and is called the  {\it slow Lasso rate}, in comparison with the {\it fast Lasso rate} 
$ s   \eps\,   \log p$ that can be obtained only if  the, so-called, {\bf compatibility }
assumption  (see, e.g., \cite{sara}) is satisfied.

In the case of the ill-posed problem \fr{geneq}, in order to achieve fast Lasso rate, 
we also need to formulate a compatibility assumption.
For this purpose,  consider a  set of $p$-dimensional vectors 
\be \label{Jset}
\calJ (\mu, J)  = \lfi \bd \in \RR^p:\ \|(\bUp \bd)_{\Jc}\|_1 \leq \mu  \|(\bUp \bd)_{J}\|_1  \rfi, \quad \mu >1,
\ee
where matrix $\bUp$ is defined in \fr{bWbUp}.
We assume that the  following condition holds:  \\

\noindent
{\bf (A)} \ Matrices $\bPhi$ and $\bUp$ are such that 
\be \label{comp}
\kappa^2 (\mu, J) = \min \lfi \bd \in \calJ(\mu, J),\, \| \bd \|_2 \neq 0: \quad 
\frac{\bd^T \bPhi \bd \cdot \Tr(\bUp_J^2)}{\|(\bUp \bd)_{J}\|_1^2} \rfi  >0.
\ee 

\noindent
Assumption  \fr{comp} is not easy to check in practice. For this reason, in the next  
section, we provide verifiable  sufficient conditions that guarantee that condition  {\bf A}  holds with  
$\kappa^2 (\mu, J)$ being uniformly bounded below  by a quantity which is separated from zero.

Observe that, in the regression setup,   $\bUp$ is the identity matrix, and  condition {\bf A} 
reduces to the compatibility condition for general sets in the Section 6.2.3  of  \cite{sara}.
If one has an orthonormal basis instead of an overcomplete dictionary, then matrix $\bPhi$ 
is an identity matrix and, due to Cauchy inequality, $\kappa^2 (\mu, J)  \geq 1$ for any $\mu$
and $J$. On the other hand, for an orthonormal basis, the bias $\| f_{\bt } - f \|_2$
in \fr{slowlas} may be large. 
Under conditions {\bf  A0 } and {\bf  A }, one obtains fast convergence rates for the Lasso estimator.

\begin{theorem} \label{th:fast_Lasso}
Let Assumptions  {\bf  A0 } and {\bf  A } hold. 
For any $\tau >0$, let  $\af = \alfo (\mu +1)/(\mu -1)$ where $\af_0$ is  defined in \fr{alf0} and $K_0 = 2$.
Then,   with probability at least $1 - 2 p^{-\tau}$,  one has
\be  \label{fasrlas}
\| f_{\hbte} - f \|_2^2   \leq   \inf_{\bt, J \subseteq \calP} \lkv  \| f_{\bt } - f \|_2^2 + 4 \al   \sum_{j \in \Jc} \nuj |t_j|    
  +   \frac{4 K_0 \mu^2 (\tau +1)}{(\mu-1)^2 \kappa^2 (\mu, J)}  \eps \log p \ \sum_{j \in J} \nu_j^2 \rkv.  
\ee 
Therefore, 
\be \label{corfast}
\| f_{\hbte} - f \|_2^2   \leq \inf_{J \subseteq \calP} \lfi \| f - f_{\calL _J} \|_2^2 + 
\frac{4 K_0 \mu^2 (\tau +1)}{(\mu-1)^2 \kappa^2 (\mu, J)}  
\eps \log p \ \sum_{j \in J} \nu_j^2 \rfi,
\ee 
where $f_{\calL _J}  = \proj_{\calL_J} f$.
\end{theorem}

Note that inequality \fr{corfast} ensures that, up to a $\log p$ factor,  the estimator $f_{\hbte}$
attains the minimum possible mean squared error for a particular function of interest $f$ as long as 
compatibility factor $\kappa (\mu, J)$ stays uniformly bounded below. 
Indeed, if $f$ were known, one would choose    $J \subseteq \calP$ and estimate $f$ by its   projection $\tilde{f}$ on $\calL_J$,
so that the overall error is  bounded below by 
\be \label{lowbound}  
\EE \|\tilde{f} - f \|_2^2 \geq  \min_{J \subset \calP} \  \lfi  \| f - f_{\calL _J} \|_2^2 +   
\frac{\eps}{\lamin(|J|, \bPhi)}\  \sum_{j \in J} \nu_j^2 \rfi,  
\ee
where $\lamin(\cdot)$ is defined in \fr{eigrestrict}. If $\kappa^2 (\mu, J)$ is bounded below
by a constant, then the lower bound in \fr{lowbound} differs from the upper bound in \fr{corfast}
by a logarithmic factor $\log p$ that serves as a price for choosing a subset of dictionary functions.

\begin{remark}\label{Las_param} {\bf (The choice of the Lasso penalty parameter)   }
{\rm 
Note that Theorems~\ref{th:slow_Lasso}~and~\ref{th:fast_Lasso} provide explicit 
expressions for the penalty parameters $\alpha$ that guarantee the slow and the fast Lasso rates.
In practice, however, those parameter values may be too high and one gets more precise estimators 
using some kind of cross validation. Another options is to set $\alpha = \hat{\al}$ where
$\hat{\al} = \arg \min_\al   \lkv  \|\bW \hbte(\af) -  \bgamma \|^2_2  + 2 \eps \hat{p} \rkv$.
Here, $\bW$ and $\bgamma$ are defined in \fr{bWbUp} and \fr{las_sol2}, respectively,  
and 
$\hat{p} = \mbox{dim} (\bW \hbte_{active})$, the dimension of the linear space $\bW \hbte_{active}$
where $\hbte_{active}$ is the reduction of  $\hbte$ to the sub-vector of the active coefficients. 
Here,  $\hat{p}$ can be viewed as the SURE estimator of the number $p(\af)$ of parameters  in the model (see \cite{tibsh}).
 }
\end{remark}

\begin{remark}\label{Dictionary_choice} {\bf (The choice of overcomplete dictionary)   }
{\rm The  choice of an overcomplete dictionary in regression problems is usually motivated by two considerations:
the dictionary should be rich enough that the function of interest allows sparse representation and also should 
satisfy compatibility conditions. In the case of the ill-posed regression problems, one has an additional constraint that 
Assumption {\bf A0} should be satisfied with $ \| \psi_j \|_{\calH_2}  < \infty$. As long as this additional constraint
holds, the issues of dictionary selection in the regression and the linear ill-posed problems are similar. 
  }
\end{remark}


\section{Discussion of the compatibility condition}
\label{sec:compat}
\setcounter{equation}{0}

Note that condition \fr{comp} is guaranteed by combination of two kinds of assumptions. 
As we have already mentioned, since the ``price'' of estimating coefficients varies from one dictionary function to the other, 
one  needs to make sure that Lasso selects coefficients with relatively low variances and sets to zero the ones with high variances.
This would be useful if the true function $f$ does not have those components. 
For this purpose, we consider the set of subsets $J \subset \calP$ such that
\be \label{Gset}
\calG (\Cups) = \lfi J \in \calP:\  \max_{j \in J,\, j' \in \Jc} \frac{ \nu_j}{ \nu_{j'}} \leq \Cups  \rfi.
\ee
We assume that the true function $f$ is such that its best approximation can be achieved using $J \in \calG (\Cups)$.
\\

\noindent
{\bf (A1)} For some $\Cups >0$ one has 
\be \label{assA1}
 \widehat{J} = \arg\min \lfi J \subset \calP:\   \| f - f_{\calL _J} \|_2^2 + \frac{4 K_0 \mu^2 (\tau +1)}{(\mu-1)^2 }\,   
\eps   \ \sum_{j \in J} \nu_j^2 \rfi \in \calG (\Cups).
\ee
Note that Assumption {\bf  A1} is natural and is similar to the usual assumptions
that $f$ is smooth and does not have  fast oscillating components. In the context of the ill-posed problems,
Assumption~{\bf  A1}  means that $f$ is not ``too hard'' to estimate.
\\

\noindent
The second  condition  needs to ensure  that   the dictionary $\lfi \ph_j,\ j \in \calP \rfi$ 
is incoherent. The latter can be warranted by one of the following  
alternative assumptions introduced in  \cite{bickel_ritov_tsybakov}.
In what follows, $\lamin$, $\lamax$ and $\varrho$ refer to   matrix $\bPhi$. 
\\

\noindent
{\bf (A2(a))} \ For some $s$, $1 \leq s \leq p/2$, some $m \geq s$ and some constant $C_0$ one has
\be \label{a2b}
\lamin (s+m) > C_0 \lamax (m),
\ee
where $\lamin (s+m)$   and $\lamax (m)$  are restricted eigenvalues  defined in \fr{eigrestrict}.
\\

\noindent
{\bf (A2(b))} \ For some $s$, $1 \leq s \leq p/2$, and some constant $C_0$ one has 
\be \label{a2c}
\varrho < [s (2 C_0  + 1)]^{-1},
\ee
where $\varrho$ is defined in \fr{nondiag}.
\\

\noindent
If Assumption   {\bf  A1} is valid, then one can replace $J \subset \calP$ by $J \in  \calG (\Cups)$
in the inequality \fr{corfast}. For $J \in  \calG (\Cups)$,   Assumption  {\bf A2(a)} (or  {\bf A2(b)})
yields a convenient lower bound on the compatibility factor $\kappa  (\mu, J)$.
In particular, small modifications of Lemma 4.1. of \cite{bickel_ritov_tsybakov} leads to the following result:

\begin{lemma} \label{lem:comp_cons} {\bf (Lemma 4.1 of \cite{bickel_ritov_tsybakov})}
Let   Assumption  {\bf  A2(a)} or {\bf  A2(b)} be valid with $C_0 = \mu \Cups$.
Then,  for any set $J \in  \calG (\Cups)$ of cardinality  $|J| \leq s$, Assumption {\bf A} holds with
$\kappa^2 (\mu, J) \geq  \vartheta (s,m)$  where
\be \label{comconst}
\vartheta (s,m) = \lfi
\begin{array}{ll}
%
\lamin(s+m) \lkr 1 - \frac{\mu \Cups \sqrt{s \lamax(m)}}{\sqrt{m \lamin(s+m)}} \rkr^2 & \mbox{if {\bf  A2(a)} holds}  \\ 
&  \\ 
1 - [s (2 \mu \Cups  + 1)]^{-1}  & \mbox{if {\bf  A2(b)} holds}.   
\end{array} \right.
\ee
\end{lemma}

\medskip

\noindent
Combination of \fr{corfast} and  \fr{comconst}  ensures  that if $f$ allows sparse representation in the dictionary 
$\lfi \ph_j, j \in \calP \rfi$, so that   set $\widehat{J}$ in Assumption {\bf  A1} has at most $s$
components, then Lasso provides an optimal (up to a logarithmic factor) representation of the   function $f$.

\begin{corollary}  \label{cor:lasso_fast}
Let Assumptions  {\bf A0}, {\bf A1} and {\bf A2(a)}  or  {\bf A2(b)}  hold with some $m$ and $C_0 = \mu \Cups$.
Let set $\widehat{J}$ in Assumption {\bf A1} have at most $s$ components: $|\widehat{J}| \leq  s$.
Then,  for  any $\tau >0$  and   $\af = \alfo (\mu +1)/(\mu -1)$, with probability at least
$1 - 2 p^{-\tau}$,  one has  
\be \label{optim}
\| f_{\hbte} - f \|_2^2   \leq \inf_{J \subseteq \calP} 
\lfi \| f - f_{\calL _J} \|^2 + \frac{4 K_0\, \mu^2 (\tau +1)}{(\mu-1)^2 }\    
\frac{\log p}{\vartheta (s,m)}\ \eps   \ \sum_{j \in J} \nu_j^2 \rfi .
\ee
\end{corollary}

\medskip

\noindent
Finally, we comment about the choice of $m$ in Assumption {\bf A2(a)}.  Similarly to regression set up, this
choice depends on how fast the  the minimal eigenvalues of the order $m$ sub-matrices of $\bPhi$ are decreasing as functions of $m$
(see, e.g., \cite{bickel_ritov_tsybakov}).

\begin{remark}\label{invertible} {\bf (Invertible dictionary matrix)   }
{\rm 
Note that if one imposes a somewhat stronger condition
\be \label{comcons2}
\max_{j'}    \sum_{j \neq j'}  |\Phi_{jj'}| \leq \kappa_0 < 1
\ee
for some $\kappa_0 >0$, then $\lamin(\bPhi) \geq 1- \kappa_0 \geq \kappa (\mu, J)$ and Assumption {\bf A} holds.
This is a ``low-dimensional'' application of Lasso technique which, however, may be of 
use in some practical situations.
}
\end{remark}


\section{Observational model}
\label{sec:obser_model}
\setcounter{equation}{0}

Consider a real-life observational model corresponding to equation \fr{geneq}
\be \label{obs}
y_i =  q(x_i) + \xi_i,\quad i=1, \cdots, n,  
\ee
where  $\xi_i$ are i.i.d. centered sub-gaussian random variables such that for some $\sigma$ and any $t$
\be \label{large_devsig}
\PP \lkr |\xi_i| > t \rkr \leq \exp(-t^2/2 \sig^2).
\ee 
Assume that $x_i \in \calX$, $i=1, \cdots, n$, are fixed non-random  points where $y(x)$ in equation \fr{geneq} is measured. 
To be more specific, we consider the case when $\calX = [a,b]$ is an interval,  
$a = x_0 < x_1 < \cdots x_n = b$  and $\calH_2 = L^2 [a,b]$, so that 
\be \label{func}
\bej = \lan   q,   \psij \ran_{\calH_2} =  \int_{\calX} q(x) \psij (x) dx.  
\ee
Denote   $T = b-a$, $\Dxi = x_i - x_{i-1}$ and define new values of $\nuj$ and $\hbej$
\be \label{bejnuj} 
\hbej  = \frac{1}{n}\ \sumin y_i \psij (x_i) \Dxi, \quad  
\nuj^2   =   \frac{T^2}{n} \ \sumin \psij^2(x_i).     
\ee 
We search for $\hbte$ as a solution of optimization problem    \fr{las_sol1} with $\hbej$ and $\nuj$ 
given by \fr{bejnuj}. 
We expect that, if  $|\Dxi|$ are small and $n$ is large enough, one can estimate 
$f$ on the basis of discrete data in  \fr{obs} as well   as on the basis of the white noise 
model \fr{geneq}. Denote
\be \label{alephdef}
\aleph   = \max_{1 \leq j \leq p} \lkv \frac{1}{\nuj}\ \max_{x \in \calX}   \left|\frac{d  [q(x) \psij(x)]}{dx}\right|\, \rkv.
\ee 
Then, the following statement holds.

\begin{theorem} \label{th:obs_Lasso}
Let Assumptions  {\bf A0} and {\bf A}  hold and $\tau>0$ be an arbitrary constant.
%
If for some non-negative constant  $\vartheta$ one has 
\be    \label{alephcond} 
\max_i |\Dxi|   \leq    \vart\, \frac{T}{n} \quad \mbox{and} \quad 
n  \geq  \calN =     \frac{T^4\, \aleph^2}{4 K_0\, \sig^2 (\tau +1) \log p}, 
\ee    
then,  for $\alfo =  2   \vart \, n^{-1/2}\, \sig\  \sqrt{2 (\tau +1) \log p}$,     
 $\af = \alfo (\mu +1)/(\mu -1)$   and  $K_0 = 8 \vart^2$, with probability at least
$1 -   e p^{-\tau}$,  inequalities \fr{fasrlas} and \fr{corfast} hold. 
\end{theorem}

 Note that the estimator $f_{\hbte}$ is fully adaptive since $\alfo$ is known. 
The lower bound $n \geq \calN$ for $n$ is motivated by the fact that the rectangular  rule approximations 
of the integrals in \fr{func} should be close in value to those integrals. 
In addition, if functions $\psi_j$ and $q$ are smooth, so that functions $q \psij$ have uniformly bounded second derivatives,  
one can replace the rectangular rule for calculating $\beta_j$ by the trapezoid rule. 
In this case, oracle inequalities in Theorem \ref{th:obs_Lasso} 
can be obtained with a smaller value of $\calN$.


\section{Lasso recovery of a mixing density from a continuous mixture}
\label{sec:mix_den}
\setcounter{equation}{0}

In this section we show that, with a small modification, the method used in the previous sections, can be applied to 
estimation of the mixing density in a continuous mixture.
 Consider the situation when one observes a random sample $Y_1, Y_2, \cdots, Y_n$ of a random variable $Y$ 
with  an unknown probability density function $q(y)$, $y \in \calY$,  of the form 
\be \label{mixden}
q(y) = \int_{\calX} g(y|x) f(x) dx,\quad y \in \calY,
\ee
where $g(y|x)$ is a known conditional density of $Y$ given $X=x$, $x \in \calX$, and $f(x)$, $x \in \calX$, 
is an unknown mixing density of interest. 
If $g(y|x) = g(y-x)$, then problem \fr{mixden} reduces to the extensively studied density deconvolution problem 
(see, e.g., \cite{meister} and references therein). In a general set up, problem  \fr{mixden} is usually solved 
by expanding $f$ over some orthonormal dictionary and then recovering coefficients of the expansion  
(see, e.g., \cite{comte},  \cite{hern} and \cite{walter}), by the kernel method (see, e.g., \cite{goutis})
or by maximizing the empirical likelihood (see, e.g., \cite{levin}). 
It is easy to see that when the conditional density $g(y|x)$ is known, the problem of recovering $f$
in \fr{mixden} on the basis of observations from $q$ can be viewed as a particular case of the linear inverse problem
\fr{obs} with the main difference that one can sample from the pdf $q$ instead of having noisy observations of the values of $q$.
Hence, one can easily estimate any linear functional of $q$, so that observations are taken in the form \fr{eq:observ}.
 For this reason, in this set up, one again can benefit from using a large overcomplete dictionary  which allows a compact representation of $f$.

Let, as before, $\lfi \ph_k  \rfi_{k=1}^p$ be a  dictionary and function $f$ be 
expanded over this dictionary yielding its  approximation \fr{f_expan}.  The goal is to recover the vector of 
coefficients $\bte$. By introducing Hilbert spaces $\calH_1 = L^2(\calX)$ and $\calH_2 = L^2(\calY)$  
and a linear operator $Q:\calH_1  \rightarrow \calH_2$ given by
\be \label{QQstarMix}
(Qf) (y) = \int_{\calX} g(y|x) f(x) dx \quad \mbox{with} \quad (Q^* u) (x) = \int_{\calY} g(y|x) u(y) dy,
\ee  
one can essentially reduce the problem \fr{mixden} to \fr{geneq}.

Note that,  despite the fact that the idea of this section seems to be similar to \cite{bunea1}, 
we consider a   different problem and apply a  completely novel approach. 
Indeed, although  in \cite{bunea1}, the authors estimated the unknown pdf  by an expansion over an overcomplete dictionary
with coefficients subsequently recovered by Lasso, they assumed that   observations from the 
density of interest are available which  makes their problem similar to the regression problem. 
On the contrary, in our case, observations from the density of interest are unavailable
which  leads  to the difficulties that are  experienced in the context of the ill-posed linear inverse problems.
Really,  though expansion \fr{f_expan} leads to $q = \sum_j \te_j u_j$ with $u_j = Q \ph_j$, 
due to contraction imposed by operator $Q$,  the system of functions $\lfi u_j,\ j \in \calP \rfi$   
does not meet compatibility condition even if $\lfi \ph_j,\ j \in \calP \rfi$ does.  
On the other hand, if one starts with an incoherent dictionary   $\lfi u_j, j \in \calP \rfi$,
the system of functions $v_j = Q^{-1} u_j$ may be totally inappropriate   for estimating $f$.

In order to apply methodology of Section \ref{sec:method}, we define 
 new values of  $\bej$,  $\hbej$ and $\nuj$ 
\be  \label{hbej3}
\bej = \EE[\psij (Y_1)], \quad
\hbej  = \frac{1}{n}\ \sumin  \psij (Y_i), \quad \nuj = \| \psij \|_{\infty}.
\ee
 We search for $\hbte$ as a solution of optimization problem    \fr{las_sol} with $\hbej$ and $\nuj$
given by \fr{hbej3}. Then, the following statement is true.

\begin{theorem} \label{th:mix_den_Lasso}
Let Assumptions {\bf A0} and {\bf A } hold. Let $\hbej$ and $\nuj$ be defined in \fr{hbej3}, $\tau$ be  any positive constant 
and $\alfo = 2\, n^{-1/2}\,  \sqrt{(\tau +1) \log p}$. Denote $\af = \alfo (\mu +1)/(\mu -1)$.  If $n \geq \calN_0 = 16/9  (\tau +1) \log p$, 
then,    with probability at least $1 - 2 p^{-\tau}$,  one has
\be    \label{mixdenlas}
\| f_{\hbte} - f \|_2^2   \leq   \inf_{\bt } \lkv  \| f_{\bt } - f \|^2 + 4 \al   \sum_{j \in \Jc} \nuj |t_j|    
  +   \frac{16 \mu^2 (\tau +1)}{(\mu-1)^2 \kappa^2 (\mu, J)}\,    \frac{\log p}{n} \ \sum_{j \in J} \nu_j^2 \rkv.  
\ee  
\end{theorem}

\medskip

\begin{remark} \label{rem1}  {\bf (Smaller penalties)   }
{\rm 
Note that $\nuj^2 = \| \psij \|^2_{\infty}$ in \fr{hbej3} can be replaced by a smaller value $\nuj^2 = \Var [\psij (Y_1)]$
which leads to a smaller overall error,  
provided the number of observations $n$ is large enough, in particular, 
\bes
n \geq \calN_1 = \max_{1 \leq j \leq p} \  \lkv \frac{16\,  (\tau +1) \log p\, \| \psij\|_{\infty}^2}{ 9\, \Var [\psij] } \rkv.
\ees
Note    that, though $\Var [\psij (Y_1)]$ is unavailable  (since $f$ is unknown),
one can easily construct an upper bound for $\nuj^2$
\be \label{nujupper}
\nuj^2 \leq  
\max_{x \in \calX} \, \lkv \int_{\calY} g(y|x) \psij^2(y) dy \rkv
\ee
or estimate $\Var [\psij (Y_1)]$ from observations.
}
\end{remark}

\begin{remark} \label{rem_den}  {\bf (Estimation by a density function)   }
{\rm 
Estimator $f_{\hbte}$ obtained as a solution of optimization problem \fr{las_sol} with $\hbej$ and $\nuj$
given by \fr{hbej3} is not necessarily a probability density function since we do not require the dictionary 
functions to be nonnegative and the weights to be such that $f_{\hbte}$ integrates to one. This, however,
can be easily accomplished in the context of Lasso estimator if one uses dictionary functions that are pdfs themselves and 
add   constraints that the coefficients are nonnegative and sum to one. Note that since we are using the weighted 
Lasso penalty, those constraints do not allow to get rid of the penalty term altogether though  the non-negativity 
 condition should make compatibility assumption {\bf (A)} weaker.  However,   pursuing this extension 
of the Lasso solution is a matter of future investigations. 
}
\end{remark}


\section{Approximate inverse images of the dictionary functions}
\label{sec:approx_images}
\setcounter{equation}{0}

 Condition {\bf A0} requires that each dictionary function $\ph_j$ allows an exact inverse image 
 $\psij$ such that $Q^* \psij = \phj$ and $\nuj = \| \psi_j\|_2 < \infty$. This may not always be true 
since functions $\psi_j$ may not be easy to construct or they may have infinite norms.
In this situation, arguments of Section \ref{sec:advantages} suggest that exact inverse images $\psi_j$ can be replaced by 
approximate ones $\psijdel$.

First, let us consider the setting of Section~\ref{sec:method} where observations are taken in the form \fr{eq:observ}
and ${\calH_1}$ and ${\calH_2}$ are spaces of square integrable functions.
Let functions $\phjdel$ be  such that $\| \phjdel - \phj\|_2  \leq \del_j$ 
and $\psijdel$ is the solution of the equation $Q^* \psijdel = \phjdel$ with 
$\nujdel = \| \psijdel \|_2    < \infty$.   Then, $\hbejdel  = \lan   y,   \psijdel \ran$,
so that 
\be \label{eq:bejdel}
\hbejdel  = \bej + \sqrt{\eps} \nujdel \eta_j + h_j, \quad h_j =  \lan q, \psijdel - \psij \ran.
\ee
 where $\eta_j$ are  standard normal variables. 
Hence, application of Lemma \ref{lem:weighted_Lasso}
with $K = \sqrt{2}$,
\be \label{apprC_h}
C_{h \del} = \max_{1 \leq j \leq p} \frac{|\lan q, \psijdel - \psij \ran|}{ \nujdel \, \sqrt{2\,\eps\, \log p\, (\tau +1)}},
\quad C_{\al \del} = \sqrt{2(\tau +1)} (1 + C_{h \del})
\ee
 and $\al_0 =  C_{\al \del} \sqrt{\eps \log p}$, for any $\tau >0$   
 and any $\af \geq \alfo$, with probability at least $1 - 2 p^{-\tau}$, yields 
\fr{slowlas}, \fr{fasrlas} and \fr{corfast} with $K_0 = 2 (1 + C_{h \del})^2$.

The practical question, however, is how can one construct the functions $\phjdel$ and $\psijdel$.
Consider operator $Q Q^*: \calH_1 \to \calH_1$ and a small parameter $\del$. Construct 
functions $\psijdel = (Q Q^* + \del I)^{-1} Q \phj$ where $I: \calH_1 \to \calH_1$ is the identity operator.
Then, one can easily check that relation \fr{eq:bejdel} holds.
The value of $\del$ can be chosen so to minimize the mean squared error of estimating 
$\bej$ by $\hbejdel$ given by
\be \label{del_er}
  \EE (\hbejdel  - \bej)^2 = \eps \nujdel^2 + [\lan q, \psijdel - \psij \ran]^2,
\ee

Although function $q$ in \fr{del_er} is unknown, one can minimize \fr{del_er} with $q$ being replaced   by an estimator.
Whenever  observations are available in the form \fr{eq:observ} or \fr{obs}, one can construct a kernel or a projection 
estimator  $\hat{q}$ of $q$ and replace $q$ by $\hat{q}$ in \fr{del_er}. In the case of recovery of a 
mixing density in a continuous mixture,  $h_j$ in
\fr{eq:bejdel} is of the form 
\bes
h_j = \EE(\hbejdel) - \bej = \lan q, \psijdel - \psij \ran = \EE[\psijdel(Y) - \psij(Y)]
\ees
and can be estimated by its sample average.


\section{Applications of the theoretical results}
\label{sec:examples}
\setcounter{equation}{0}

In this section we consider two applications of the theory above. 
In order to show capabilities of the Lasso technique, in Section~\ref{sec:wishart}, we study  
  estimation of the unknown density function of the matrix parameter of the Wishart distribution.
This type of problems   is very hard to handle by traditional methods due to the curse of dimensionality.
The second example, presented in Section~\ref{sec:Laplace}, deals with the solution 
of a noisy version of the Laplace  convolution equation that appears in many practical applications. 
After theoretical treatment of the problem in Section~\ref{sec:Laplace}, we  study 
it further by numerical simulations in Section~\ref{sec:simulation}.


\subsection{Estimation of the density of the matrix parameter of the Wishart distribution}
\label{sec:wishart}

Let $\bY|\bX \sim\ {\mbox Wishart} (m, \bX)$,  where $\bX, \bY \in \RR^{r \times r}$ 
are symmetric positive definite $r$-dimensional matrices:
\be \label{wishart}
g(\bY | \bX) = \frac{|\bY|^{\frac{m-r-1}{2}} 2^{-\frac{mr}{2}}} {|\bX|^{\frac{m}{2}} \Gamma_r \lkr \frac{m}{2} \rkr } \ 
\exp \lfi - \frac{1}{2} \Tr (\bX^{-1} \bY ) \rfi, \quad \bX, \bY >0, m > 3r,
\ee
where $\Gamma_r \lkr a \rkr$ is the  multivariate gamma function  (see, e.g., \cite{gupta}, Section 1.4)
\be \label{multgam}
\Gamma_r (a) = \pi^{\frac{r(r-1)}{4}} \ \prod_{l=1}^r \Gamma \lkr   \frac{2a - l+ 1}{2} \rkr. 
\ee

Consider the situation when, given $\bX_i=\bX$, matrix $\bY_i$ has the Wishart pdf of the form 
\fr{wishart},  $i=1, \cdots, n$, and matrices $\bX_1, \cdots,  \bX_n$ are independent with the common 
unknown pdf $f(\bX)$. Here, matrices $\bY_1, \cdots, \bY_n$ are available for observation 
but $\bX_1, \cdots,  \bX_n$ are not. The objective is to  estimate the pdf  $f(\bX)$ of the unknown matrix parameter $\bX$
on the basis of   observations $\bY_1, \cdots, \bY_n$ of $\bY$. 
This problem appears, for example,  when one has several  equal size samples from the multivariate normal distribution with different 
unknown covariance matrices $\bX_1, \cdots,  \bX_n$ that are related by a common pdf $f(\bX)$. 
The estimator $\hf(\bX)$ of $f(\bX)$ can be used, for example, as a prior distribution in 
subsequent Bayesian inference.

It is a well known fact that, even for moderate values of $r$,  an estimator will suffer from the curse of dimensionality.
In order to circumvent this difficulty, we estimate $f(\bX)$ using an overcomplete dictionary.
In this example, $\calX = \calY$ are the  spaces of the symmetric 
nonnegative definite matrices in $\RR^{r \times r}$ and   $\calH_1 = \calH_2$ 
are the Hilbert spaces of square integrable functions on $\calX = \calY$. We choose a dictionary that consists of a collection of 
 mixtures of inverse Wishart densities since this is a wide class, so that, the true density $f(\bX)$ either belongs 
to this class or is well approximated by it.
In particular, we choose the dictionary functions of the form 
\bes
 \ph_j (\bX) = C_{\bA_j}\ u(\bX| \bA_j, \ga_j),\quad  j=1, \cdots, p, 
\ees
with  $2r < \ga_j < m-r$, where 
$u(\bX| \bA, \ga)$ is the inverse Wishart density and $C_{\bA}$ is the normalizing constant, such that
$u(\bX| \bA, \ga)$ has   
 the unit $L^2$-norm:  
\beqn\ph (\bX) & = &  \ph(\bX| \bA, \ga) = C_{\bA} \, u(\bX| \bA, \ga) 
\quad \mbox{with} \quad \|\ph(\bX)\|_2  =1, \nonumber\\
\label{inv_Wish}
u(\bX| \bA, \ga) & = & \frac{2^{-\frac{(\ga - r -1)r}{2}} | \bA |^\frac{\ga - r -1}{2}}
{\Gamma_r \lkr \frac{\ga - r -1}{2} \rkr |\bX|^\frac{\ga}{2}} \ 
\exp \lfi - \frac{1}{2} \Tr (\bX^{-1} \bA ) \rfi, \ \  \bX, \bA >0. 
\eeqn
By direct calculations it is easy to check that
\beqn 
C_{\bA} & = & \Gamma_r \lkr \frac{\ga - r -1}{2} \rkr\,  \lkv \Gamma_r \lkr \frac{2 \ga - r -1}{2} \rkr \rkv^{-1/2}\ 
2^{-\frac{r(\ga - r -1)}{2}}\ |\bA|^{\frac{r+1}{4}},  \label{CA}
\eeqn 
and that matrix $\bPhi$ in \fr{bWbUp}  has components  of the forms
\be \label{eq:bPhiWish}
\Phi_{i,j} = \lan \ph_i, \ph_j \ran =   \frac{ \Gamma_r \lkr \frac{\ga_i + \ga_j - r -1}{2} \rkr \, 2^{\frac{r(\ga_i + \ga_j   - r -1)}{2}}     }
{\lkv \Gamma_r \lkr \frac{2 \ga_i   - r -1}{2} \rkr\, \Gamma_r \lkr \frac{2 \ga_j   - r -1}{2} \rkr \rkv^{1/2}}\ 
\frac{|\bA_i|^{\frac{2 \ga_i   - r -1}{4}}\, |\bA_j|^{\frac{2 \ga_j   - r -1}{4}}}
{\left|  \bA_i  +  \bA_j  \right|^{\frac{\ga_i + \ga_j   - r -1}{2}}}
\ee

Functions $\psi_j (\bY)$ in \fr{hbej3} are solutions of equations $Q^* \psi_j = \ph_j$ 
where operator $Q^*$ is defined in \fr{QQstarMix}. It is easy to 
 verify that functions  $\psi_j (\bY)$ are of the forms 
\be  \label{psij_wishart}
\psi_j (\bY) = \psi  (\bY|\bA_j, \ga_j ) = C_{\bA_j}\ v(\bY| \bA_j, \ga_j), \quad  j=1, \cdots, p,
\ee
where $C_{\bA}$ is defined in \fr{CA} and $v(\bY| \bA, \ga)$ is the solution of the 
equation 
\bes
\int_{\calY} g(\bY | \bX) v(\bY| \bA, \ga) d\bY = u(\bX| \bA, \ga).
\ees
Here $g(\bY | \bX)$ and $u(\bX| \bA, \ga)$ are defined by, respectively, formulae \fr{wishart} and \fr{inv_Wish}, 
and the integral is calculated over the space $\calY$ of all  ($r \times r$) symmetric non-negative definite matrices.
By straightforward calculus, derive that 
\be \label{psiAgam}
\psi (\bY| \bA, \ga) = \frac{\Gamma_r \lkr \frac{m}{2} \rkr\  2^{\frac{ \ga  r}{2}} |\bA|^{\frac{2\ga-r-1}{4}} }
{\Gamma_r \lkr \frac{m-\ga}{2} \rkr\ \sqrt{\Gamma_r \lkr \frac{2\ga - r-1}{2} \rkr}}\ 
\frac{|\bY - \bA|^{\frac{m -\ga -r-1}{2}}   }
{|\bY|^{\frac{m-r-1}{2}} } \ \II(\bY - \bA >0).
\ee
Then, Theorem~\ref{th:mix_den_Lasso} yields the following corollary.

\begin{corollary}  \label{cor:wishart}
Let Assumption {\bf (A)} hold with matrix $\bPhi$ defined in \fr{eq:bPhiWish} and 
\be \label{nuj_wishart}
\nuj = \| \psi_{\bA_j, \ga_j} \|_{\infty} = 
\frac{ \Gamma_r \lkr \frac{m}{2} \rkr \, (m - \ga_j - r -1)^{\frac{r(m - \ga_j - r -1)}{2}}\, (2 \ga_j)^{\frac{r \ga_j}{2}}  }
{ \Gamma_r \lkr \frac{m-\ga_j}{2} \rkr\ \sqrt{\Gamma_r \lkr \frac{2\ga_j - r-1}{2} \rkr}\ 
(m - r -1)^{\frac{r(m-r-1)}{2}} }\ |\bA_j|^{-\frac{r+1}{4}}.
\ee
Let $\hbej$   be given by \fr{hbej3}, $\tau$ be  any positive constant, 
 $\alfo = 2\, n^{-1/2}\,  \sqrt{(\tau +1) \log p}$ and $\af = \alfo (\mu +1)/(\mu -1)$.  
If $n \geq \calN_0 = 16/9  (\tau +1) \log p$, 
then,    with probability at least $1 - 2 p^{-\tau}$,  inequality \fr{mixdenlas} holds.
\end{corollary}


\subsection{Solution of a noisy version of the Laplace deconvolution equation}
\label{sec:Laplace}

Consider Laplace deconvolution problem where one is interested in estimating an unknown function 
$f(z)$ on the basis of noisy measurements  $y_i =  q(x_i) + \sig \eta_i$, $i=1, \cdots, n$, 
of $q(x)$ where  
\be \label{lapl} 
q(x) = \int_0^x g(x-t) f(t) dt,  \quad  0 \leq x  < \infty,
\ee
function $g$ is assumed to be known, $\eta_i$ are i.i.d. standard normal variables 
and observations are available for $0 \leq x_i \leq T$ only.  
Equation \fr{lapl} is the, so called, Laplace  convolution equation 
and it appears in many practical applications (see, e.g., \cite{APR} or \cite{gripenberg} 
and references therein).

Fourier transform cannot be efficiently applied to solution of the noisy discrete version of equation \fr{lapl}.
Indeed, discrete Fourier transform  does not convert the right hand side of \fr{lapl}
into the product since the integral in formula \fr{lapl} does not realize circular convolution. 
Although one can apply the Fourier transform on the real line to equation \fr{lapl}, this 
application runs into multiple obstacles: for small values of $n$ and $T$, inverse Fourier transform 
has poor precision since Fourier transform inherently operates on the whole real line 
and requires integration of highly oscillatory functions.

Exact solution of \fr{lapl} can be obtained by using
Laplace transform.  However, direct application of Laplace transform on the basis of
discrete measurements faces serious conceptual and numerical problems.
The inverse Laplace transform is usually  found by application of tables of inverse Laplace
transforms, partial fraction decomposition or
series expansion (see, e.g., \cite{handbook}), neither of which is
applicable in the case of a discrete noisy version of Laplace deconvolution.
Since  the approach of the paper is based on inverting integral operators for completely known functions,
it appears to be  particularly useful in this situation.

Note that \fr{lapl} implies that $\calH_1 =  \calH_2 = L^2_{[0, T]}$ and one has
\be \label{operconj}
(Q^* u) (z)   = \int_z^T g(x-z) u(x) dx, \quad 0 \leq z \leq T. 
\ee
Since the right hand sides  of equations $Q^* \psi_j  = \ph_j$
are known exactly,   solutions $\psi_j$ with $\psi_j (x) =0$ when $x<0$,
can be obtained by using Laplace transform or any other suitable technique. 
Indeed, by introducing new functions
$\tpsi (x) = \psi(T-x)$ and $\tphi (x) = \ph(T-x)$, one can transform equation 
 $Q^* \psi  = \ph$ into equation
\be \label{exact_psi}
\int_0^x g(x-z) \tpsi (z) dz = \tphi (x), \quad 0<x<T, 
\ee 
that can be solved   by using the Laplace transform.

It turns out that, for any $a>0$, the   Laguerre functions 
\be \label{lag_fun}  
L_k(t;a) = \sqrt{2a} e^{-at} \sum_{j=0}^k (-1)^j {k \choose j} \frac{(2 a t)^j}{j!},\ \ 
k=0,1, \ldots,\ t \geq 0,
\ee 
 form a basis, which is particularly suitable for the problem at hand since 
it acts as a surrogate eigenfunction  basis for the problem (see, e.g., \cite{CPR} and  \cite{weeks}).
Functions $L_k(t;a)$ form an orthonormal basis of $L^2(0,\infty)$ space but are highly oscillatory
when $k$ is large. 
In order to accommodate different values of $a$ in   expression \fr{lag_fun}  and  
use simple dictionary functions, we choose the following collection of dictionary elements
$\calD = \lfi \ph_j \equiv \ph_{l_j, b_j}, \ j \in \calP \rfi$ with $b_j >0$ and nonnegative integer $l_j$ where   
\be \label{phikb}
\ph_{l,b} (z) =   e^{-bz}\ \frac{z^{l}\,(2b)^{l+1/2}}{\sqrt{(2l)!}}\quad \mbox{with}  \quad \| \ph_{l,b} \|_2 = 1.
\ee
Then, functions $\psij = \psi_{l_j, b_j}, \ j \in \calP$, in Assumption {\bf A0} can be obtained by either 
solving equation \fr{exact_psi} with $\tphi (x)  = \ph_{l_j, b_j} (T-x)$
or by numerical solution of equations  $Q^* \psij  = \phj$, $j=1, \cdots,p$. 
In our   simulation study we used the latter option.  

It is easy to see that $\nuj = \|\psi_{l_j,b_j} \|_2 < \infty$, so one can carry out Lasso estimation provided
the dictionary $\calD$ satisfies one of the Assumptions, {\bf A2(a)} or {\bf A2(b)}.
Specifically, the following Lemma provides  simple  upper bounds for the non-diagonal elements of the matrix $\bPhi$
and, hence, allows to choose the collection $\lfi l_j, b_j, \ j=1, \cdots, p \rfi$ in formula \fr{phikb}
such that  condition  {\bf A2(b)} is valid.

\begin{lemma} \label{lem:coherence}
Let  $\ro (l_1, l_2; b_1, b_2) = \lan \ph_{l_1, b_1}; \ph_{l_2, b_2} \ran$ be elements of matrix $\bPhi$.
Then, for any pair of indices $j, k \in \calP$ such that $l_j \leq l_k$ provided $b_j \geq b_k$,  one has
\be \label{corineq}
0< \ro (l_j, l_k; b_j, b_k) \leq \exp \lfi -  \frac{(2 l_k   +1)}{2} \lkv\, \left|\, \log \lkr \frac{b_j}{b_k} \rkr  \right| - \log 4 \, \rkv \rfi. 
\ee 
\end{lemma}


\section{Simulation study} 
\label{sec:simulation}
\setcounter{equation}{0}

\begin{figure} [h] 
\[\includegraphics[height=4.0cm]{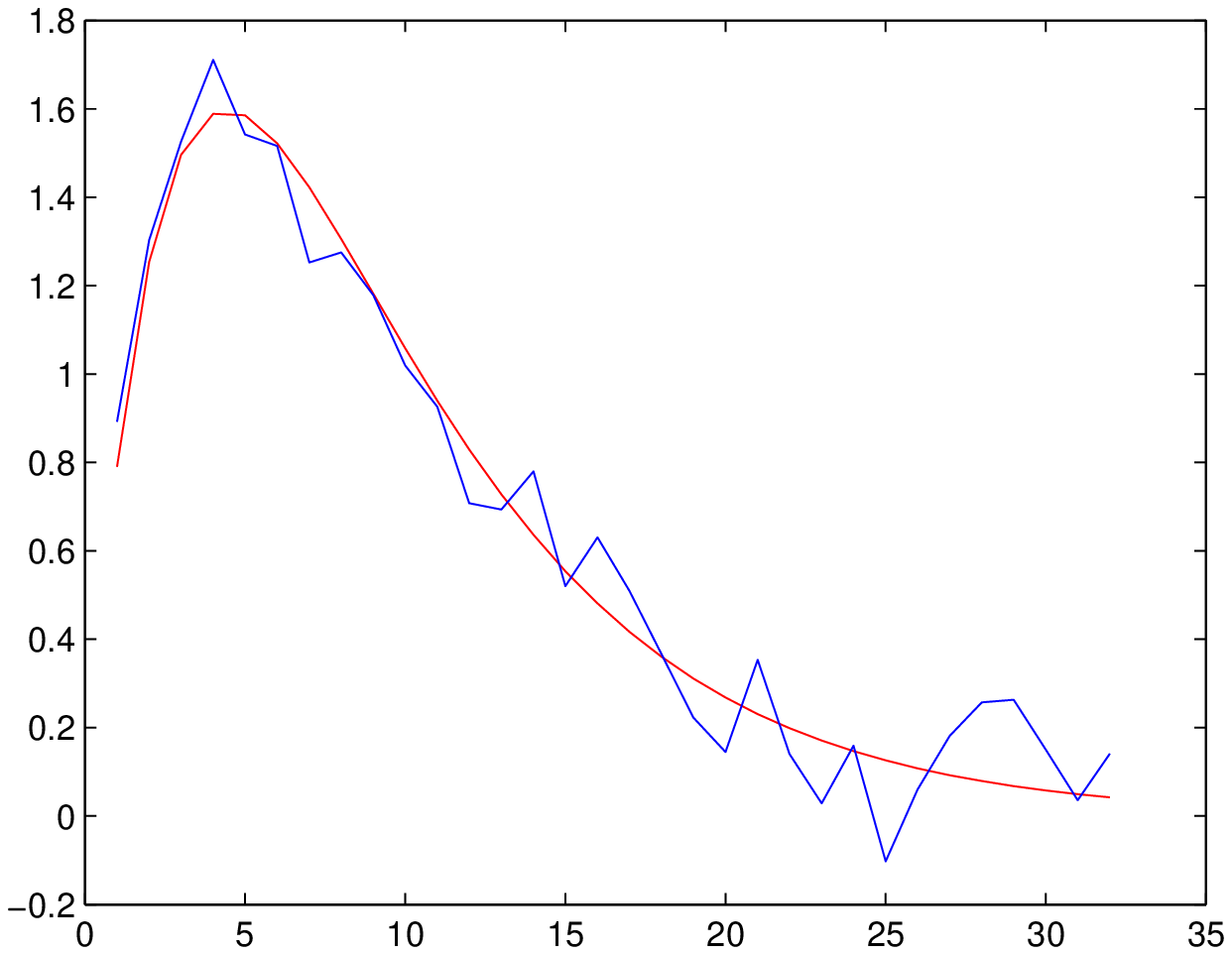} \hspace{2mm}  \includegraphics[height=4.0cm]{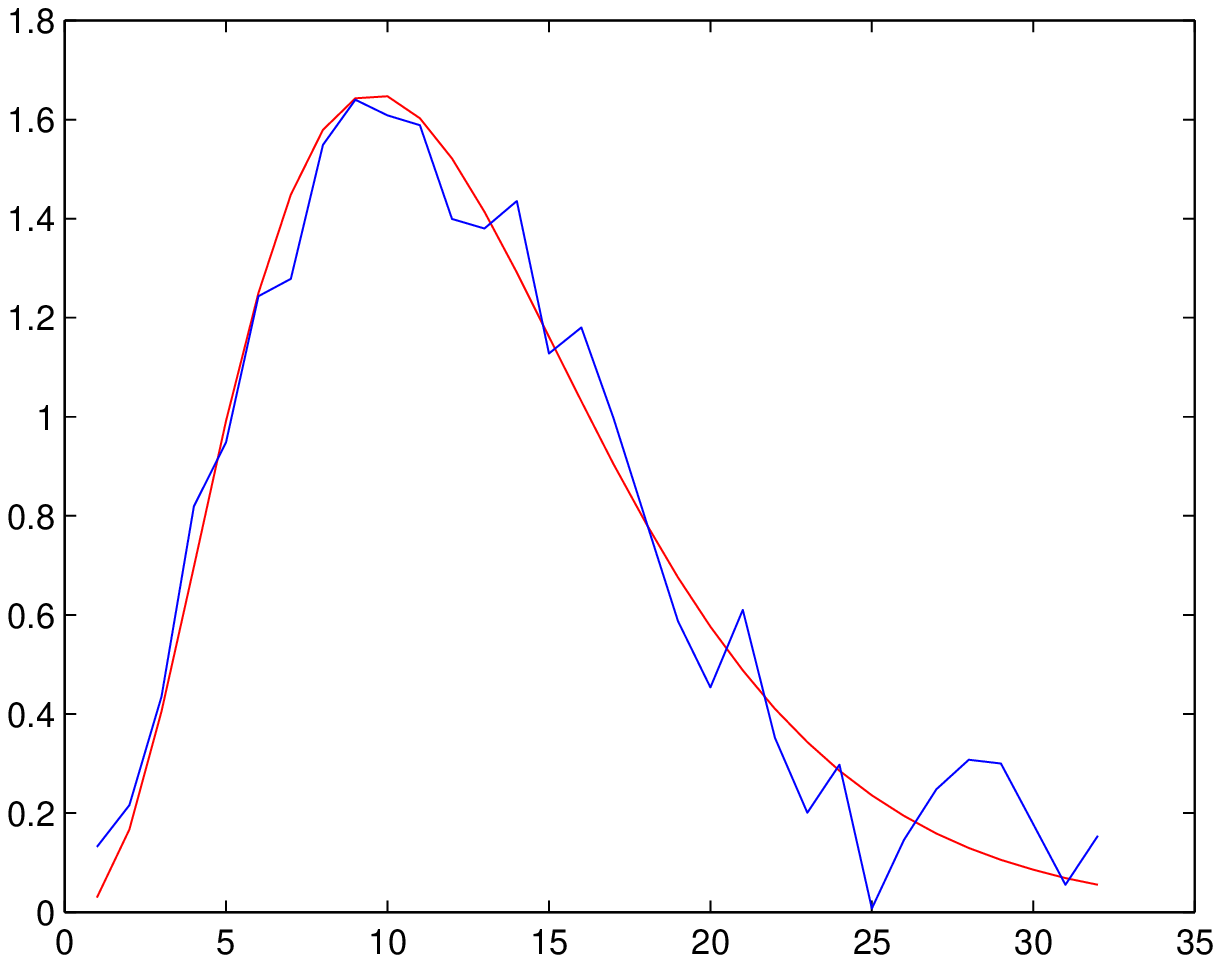}
 \hspace{2mm} \includegraphics[height=4.0cm]{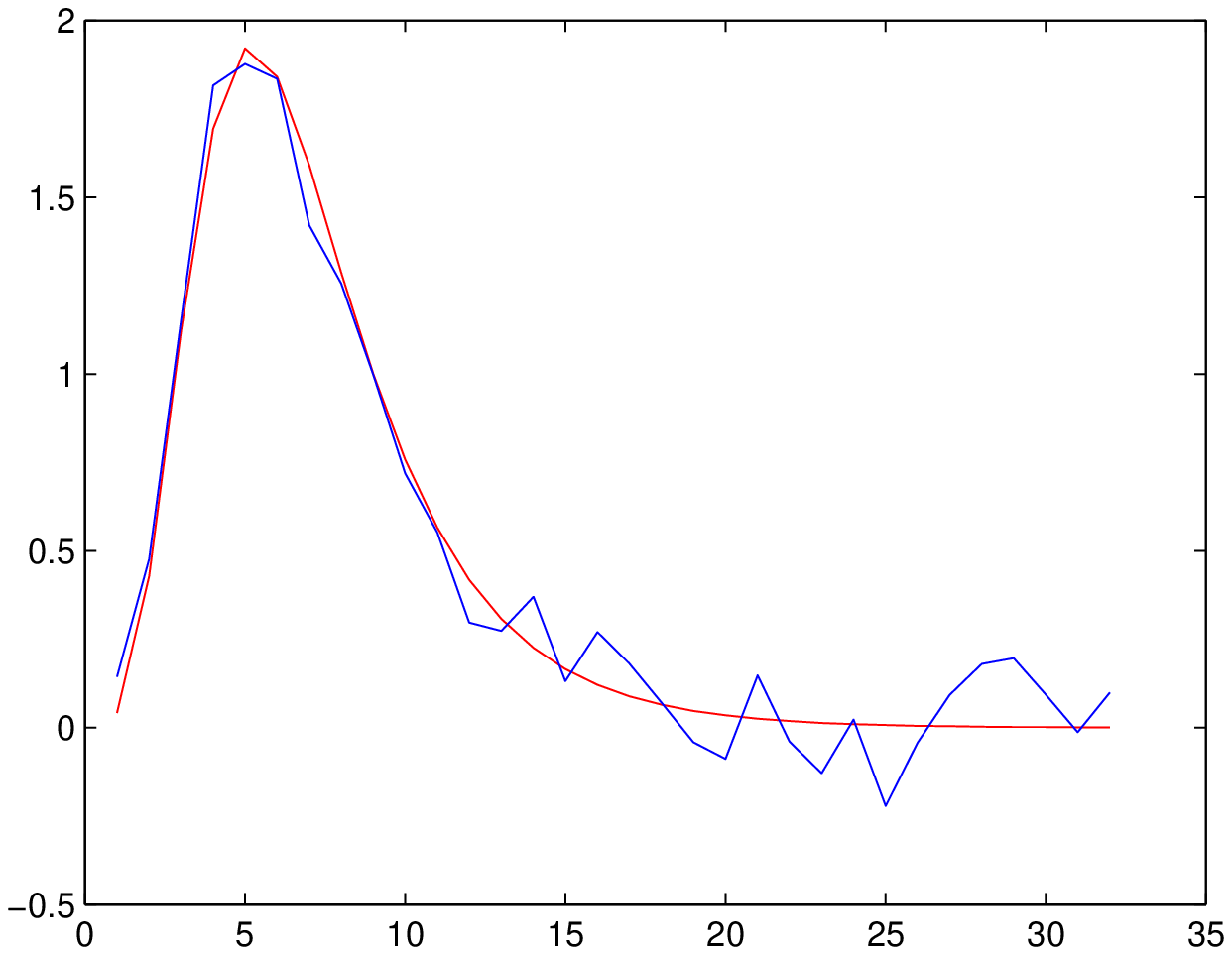} \]
\[\includegraphics[height=4.0cm]{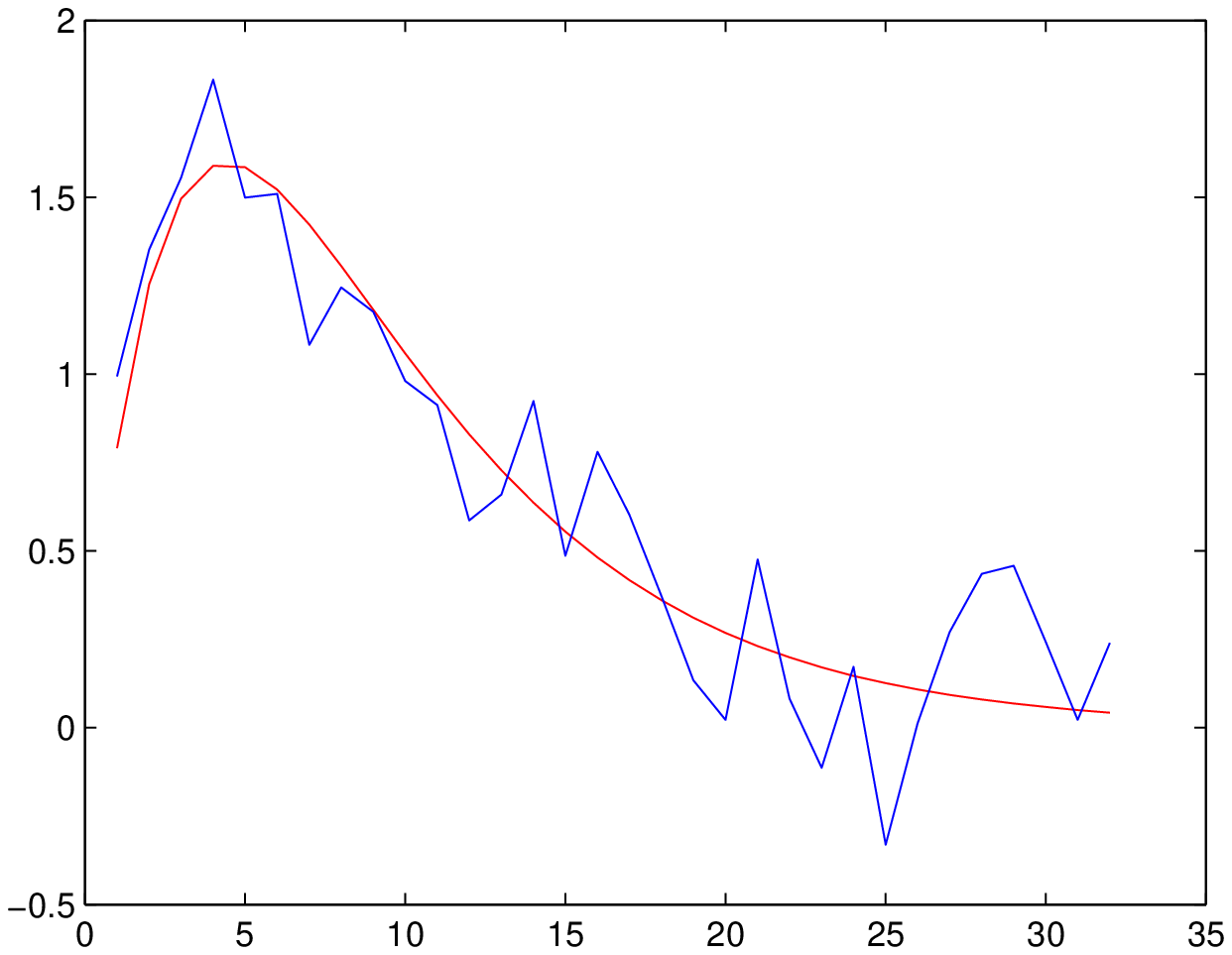} \hspace{2mm}  \includegraphics[height=4.0cm]{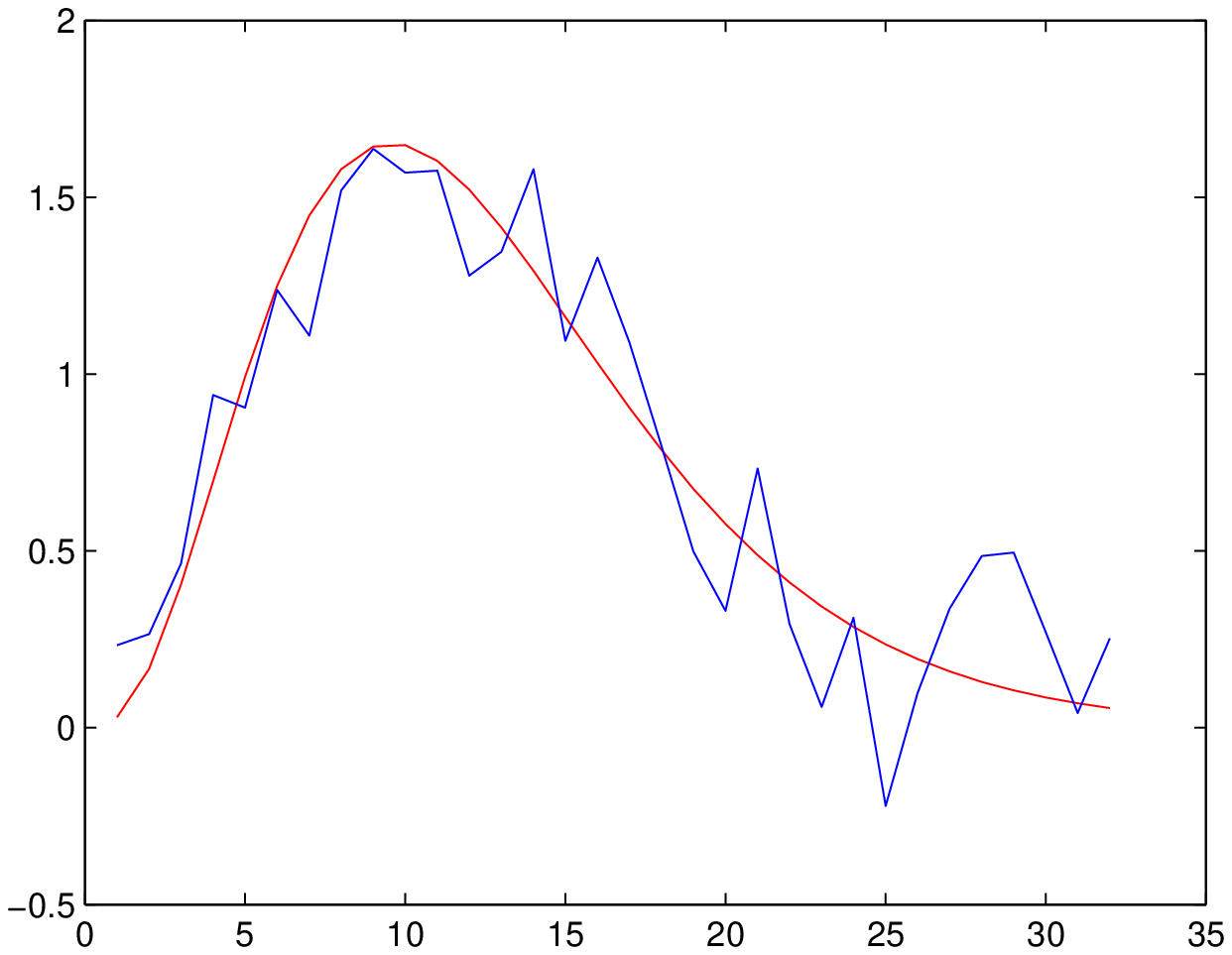}
 \hspace{2mm} \includegraphics[height=4.0cm]{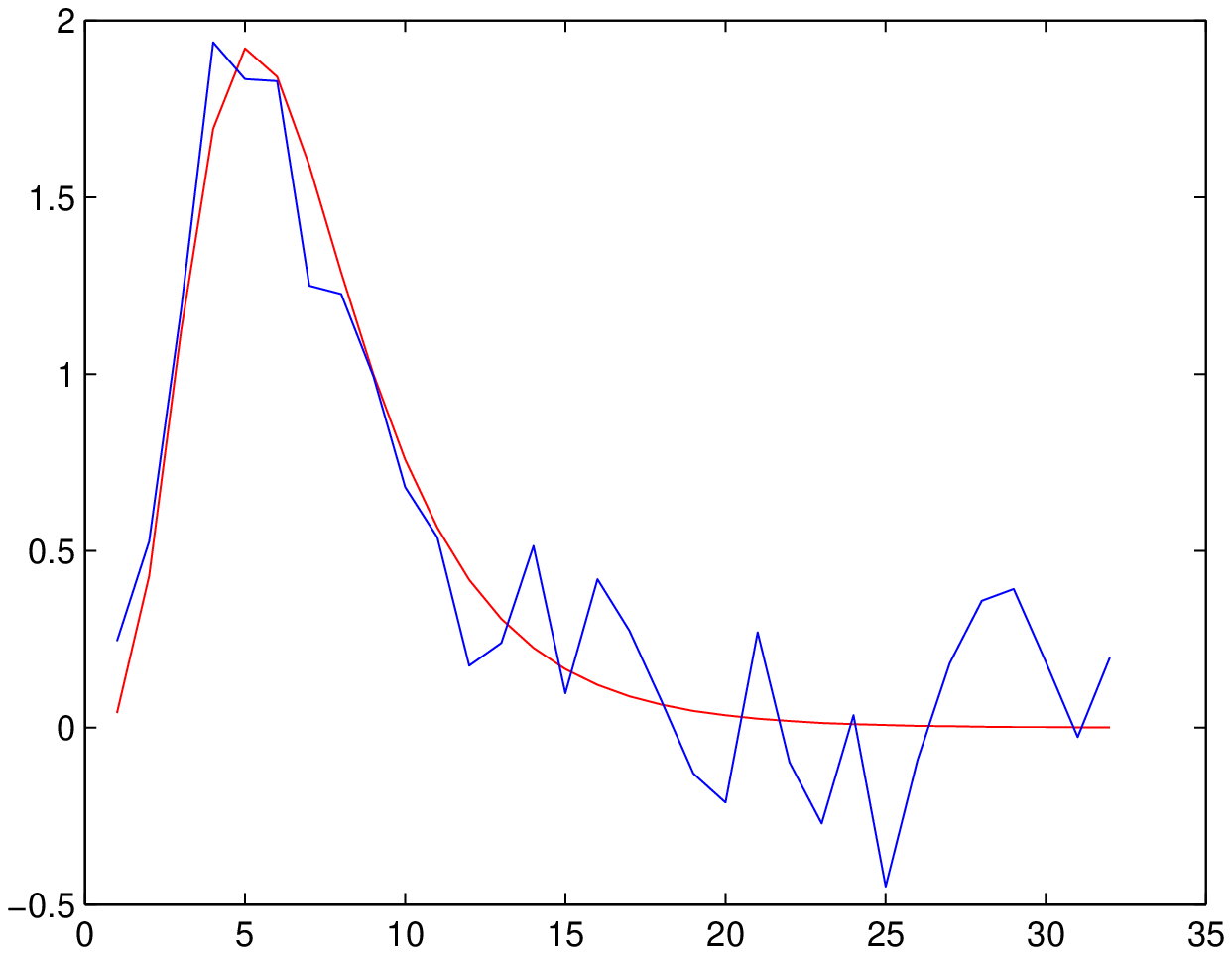} \]
\caption{The true function $q$ (red) and vector $\by$ (blue)  for test functions $f_1$ (left), $f_2$ (middle) and 
$f_3$ (right) with $n = 32$ and $\sigma = 0.5$ (top)  and $\sigma = 1$ (bottom).
\label{fig:rhs}}
\end{figure}


In order to evaluate the performance of the procedure suggested in the paper we carried out a limited 
simulation study. We considered a Laplace convolution equation \fr{lapl} studied above in Section \ref{sec:Laplace}  
with $g(x) = \exp(-x)$, observation  points uniformly spaced on the interval $[0;T]$ with $T=10$ and $n=32$ or $n=64$ observation points.  
The choice of the kernel $g$ and the sample size are motivated by the fact that Laplace convolution equation 
with the kernel of this form satisfies the conditions required for  application of 
the wavelet-vaguelette estimator which we use for comparison with our estimator.

We constructed  a fixed dictionary  of the form $\calD = \lfi \ph_j \equiv \ph_{l_{k_j}, b_{i_j}},\ j=1, \cdots, p \rfi,$   
where $\ph_{l,b}$ are defined in formula \fr{phikb} with $l_k = k$, $k = 0 \cdots p_1$, and $b_k = 0.1\  k$, 
$k=1, \cdots, p_2$.  We chose $p_1 = 10$, $p_2 = 40$, giving the total of $p = p_1 p_2 =400$
dictionary elements. We evaluated  the dictionary functions   on a fine grid, 
scaled them to have unit norms and formed matrix $\bW$   with columns $\bph_j$, $j=1,\cdots,p$.
Vector  $\by$ was calculated at $n$ observation points as $\by = \bq + n^{-1/2}\, \sig \bxi$ 
where $\bxi \in \RR^n$ is a standard normal vector.

In order to generate operator $\bQ$ we sampled functions $f$ and $g$ on a fine grid.
Matrix $\bQ$ was constructed so that it carried out numerical integration in formula 
\fr{lapl} for $0 \leq x \leq T$, i.e., $q =  Q f$.    
We obtained matrix $\bPsi$ of the inverse images by the numerical  solution of the exact equation $\bQ^* \bPsi = \bW$.
We estimated vector $\bobeta$ with elements \fr{mainrel}    by  $\hat{\bobeta} = \bPsi^T \by$
and solve optimization problem \fr{las_sol2}.   
For implementation of minimization in \fr{las_sol2}, we used function {\tt LassoWeighted} in  SPAMS MatLab toolbox \cite{spams}.
We calculated $\alpha_{\max}$ as the value of the Lasso parameter that guarantees that all coefficients in the model vanish.
We created a grid of the values of $\alpha_k = \alpha_{\max}*k/N$, $i=1, \cdots, N$, with $N = 200$. As a result, 
we obtained a collection of estimators $\hbte = \hbte(\af_k)$. Finally, we  chose $\af = \af_{opt} = \af_{\tk}$ where 
where $\tk = \arg\min_k \|\hbf(\af_{k}) - \bof ||$. 
\begin{figure}  [h] 
\[\includegraphics[height=4.0cm]{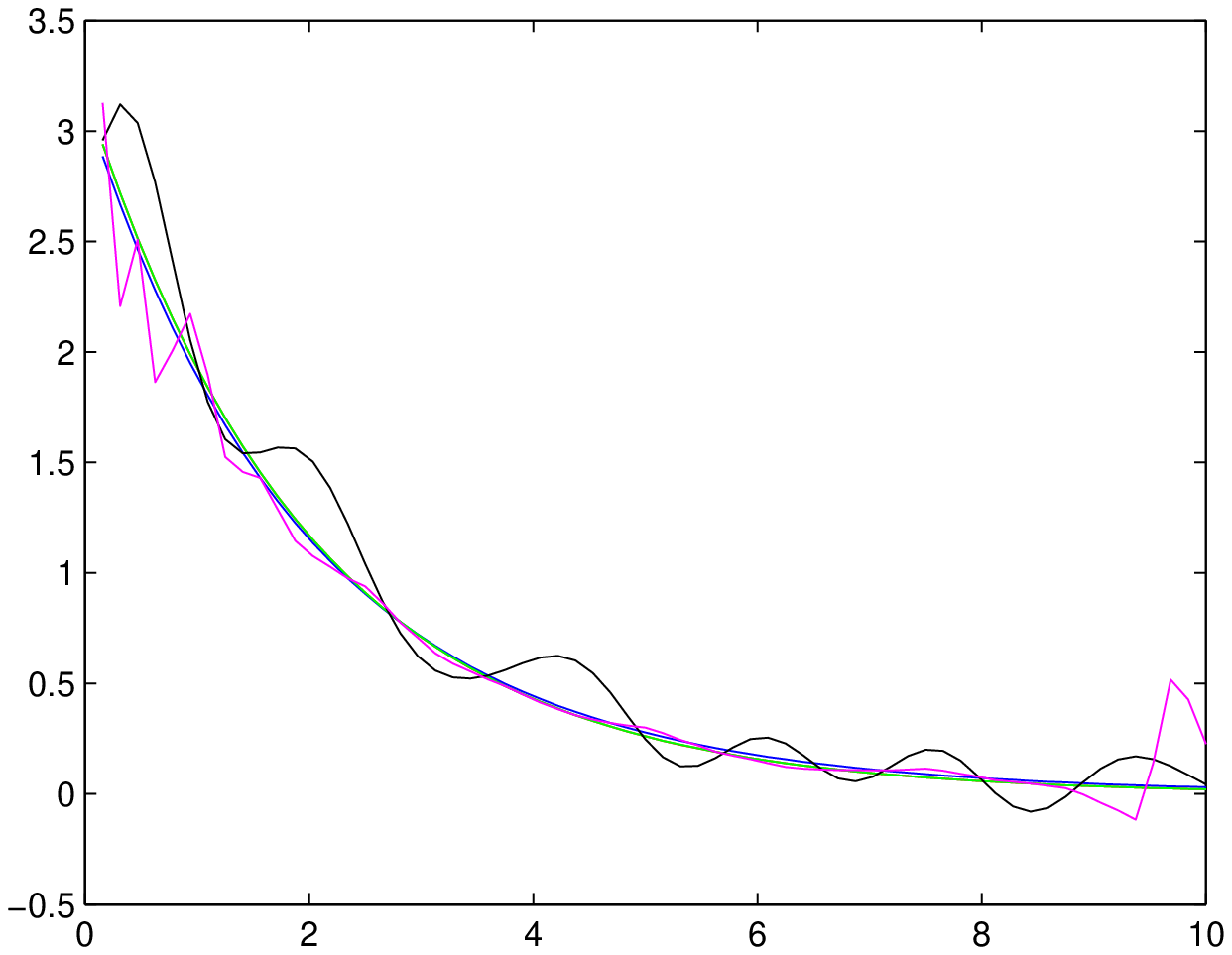} \hspace{2mm}  \includegraphics[height=4.0cm]{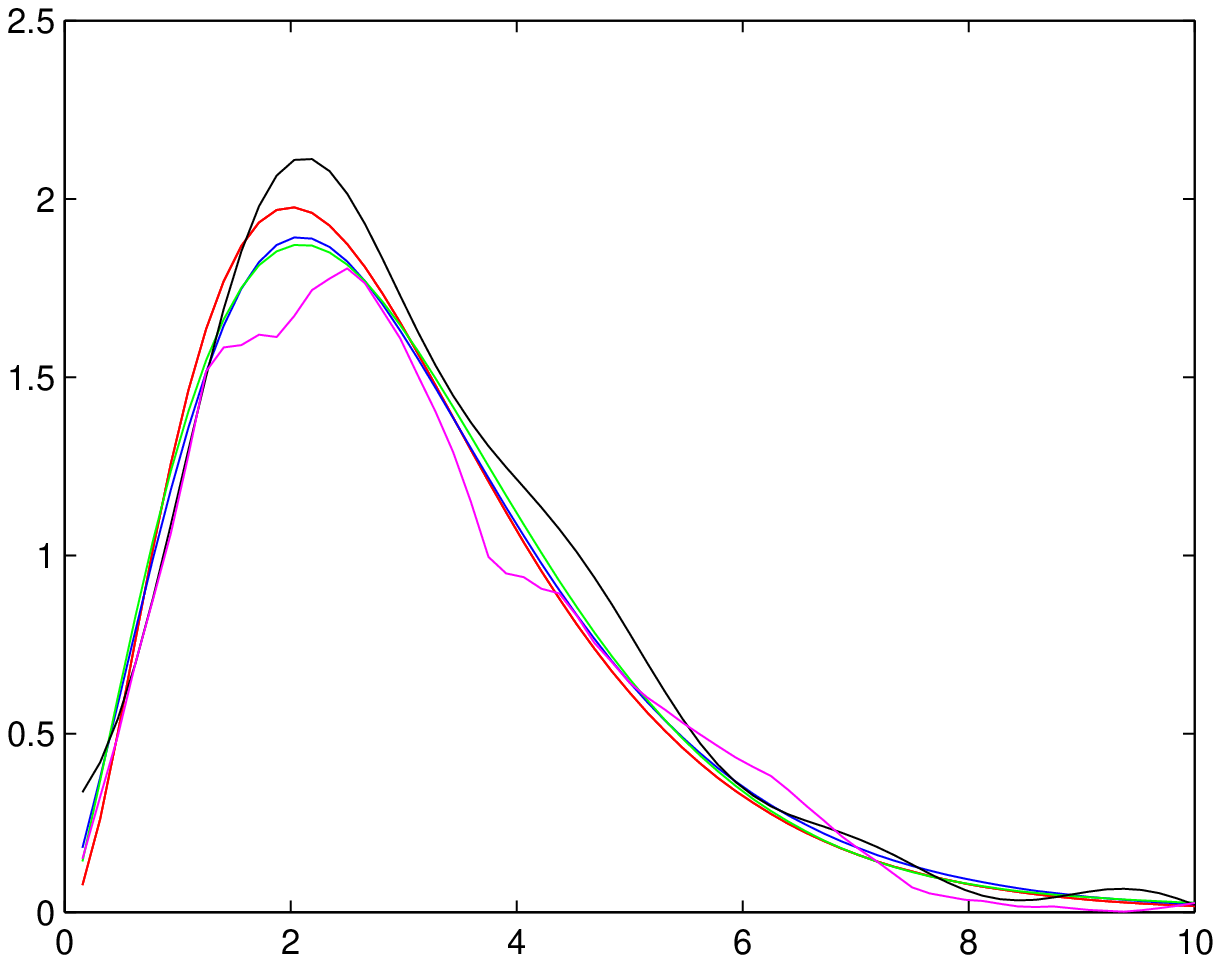}
 \hspace{2mm} \includegraphics[height=4.0cm]{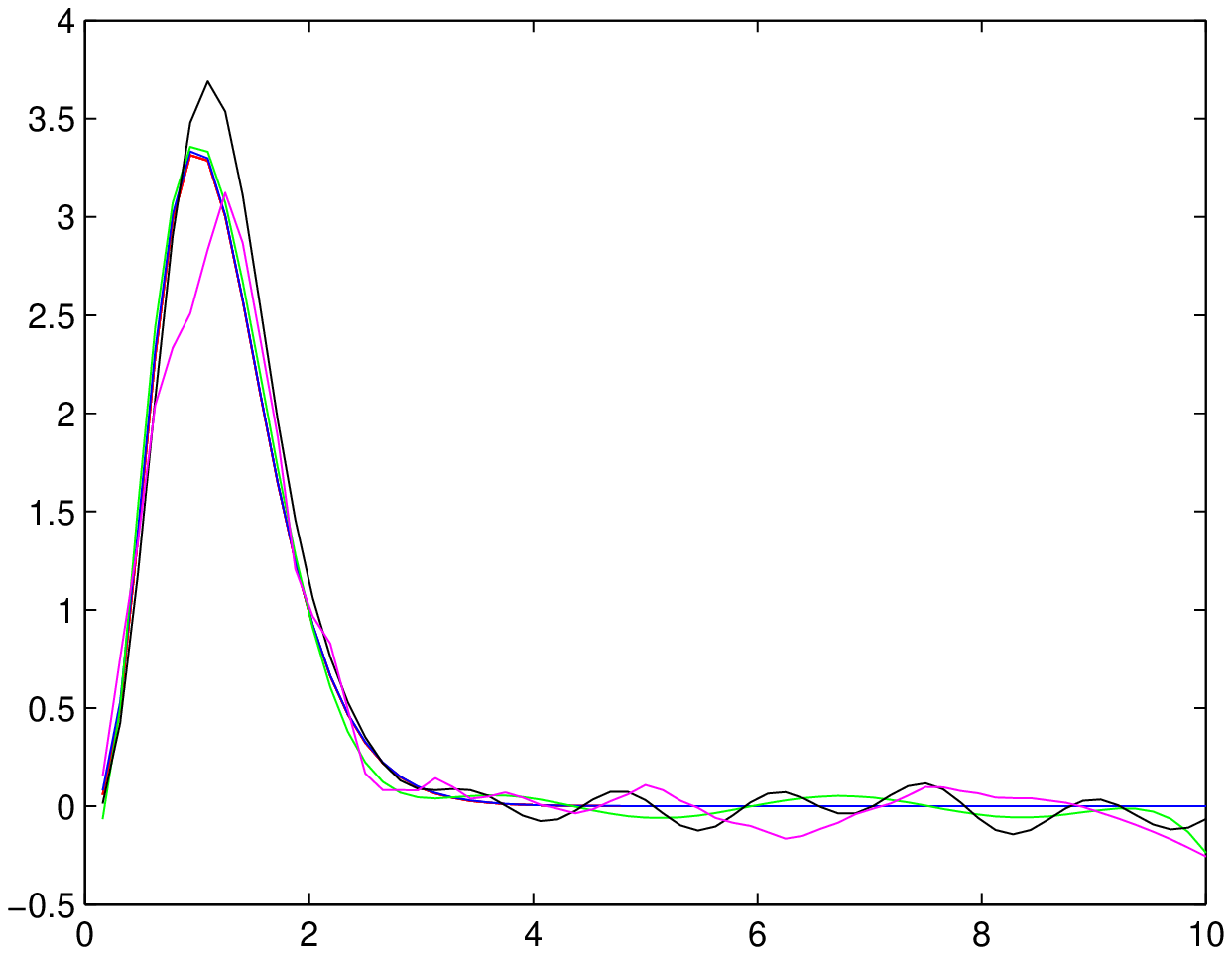} \]
\[\includegraphics[height=4.0cm]{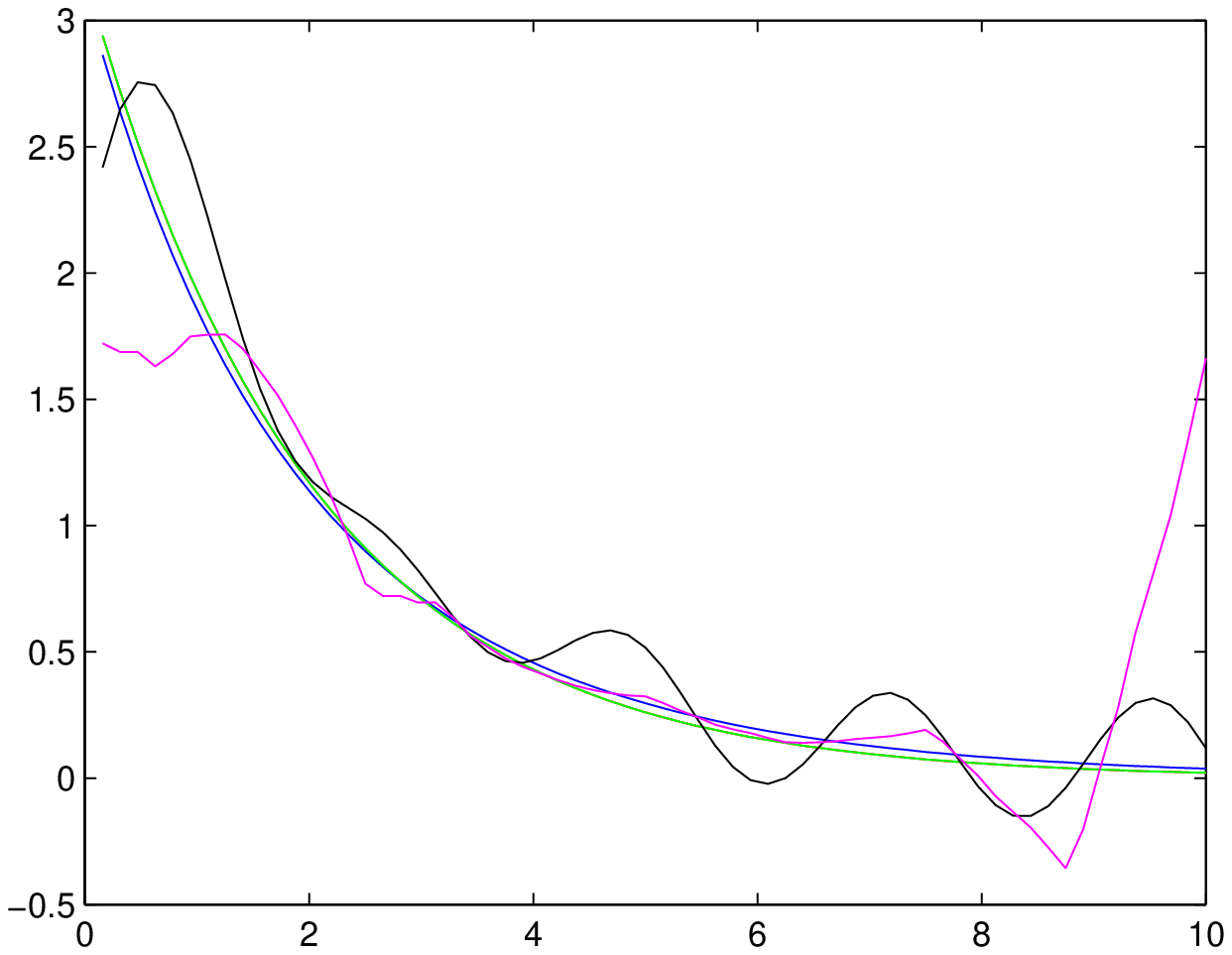} \hspace{2mm}  \includegraphics[height=4.0cm]{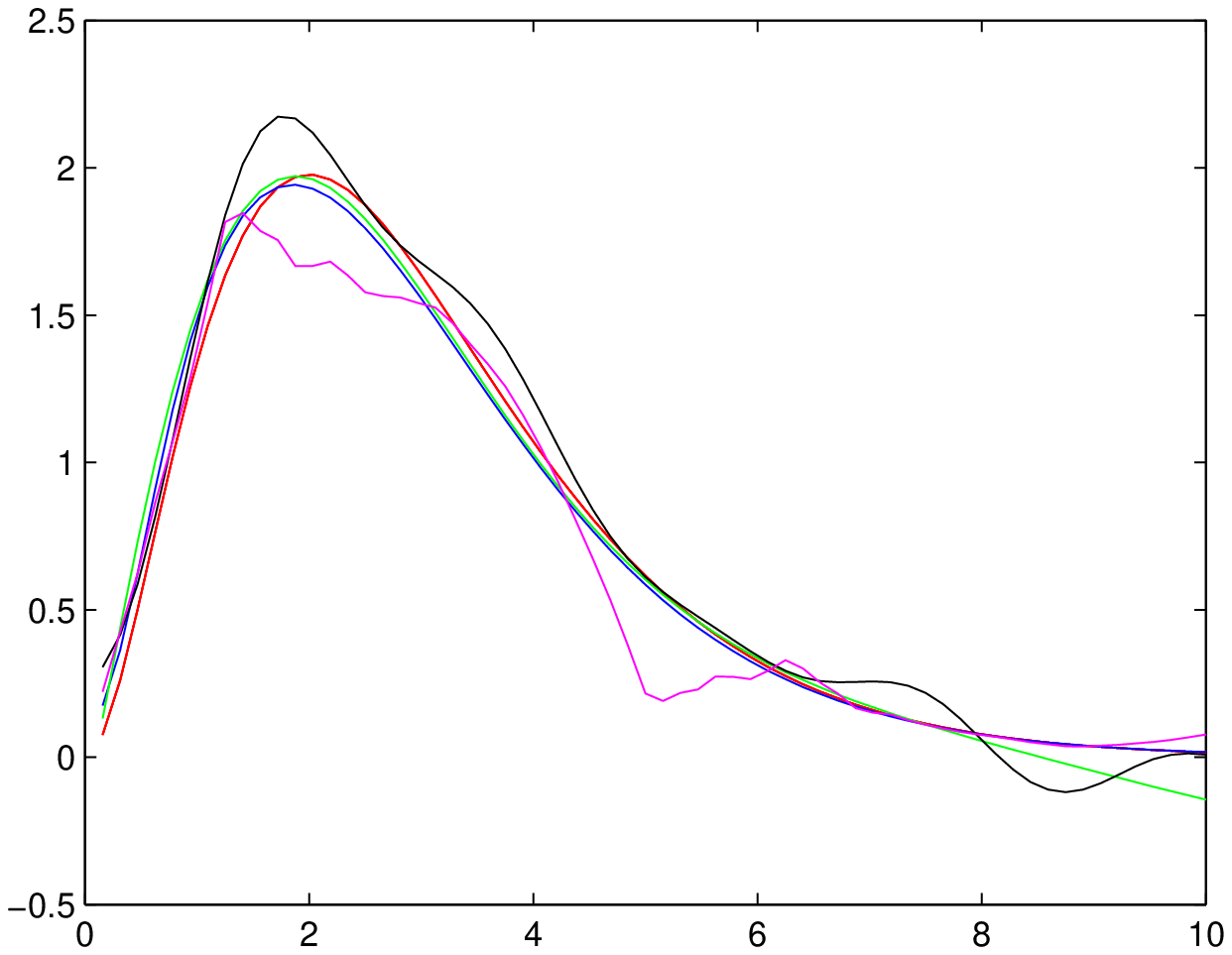}
 \hspace{2mm} \includegraphics[height=4.0cm]{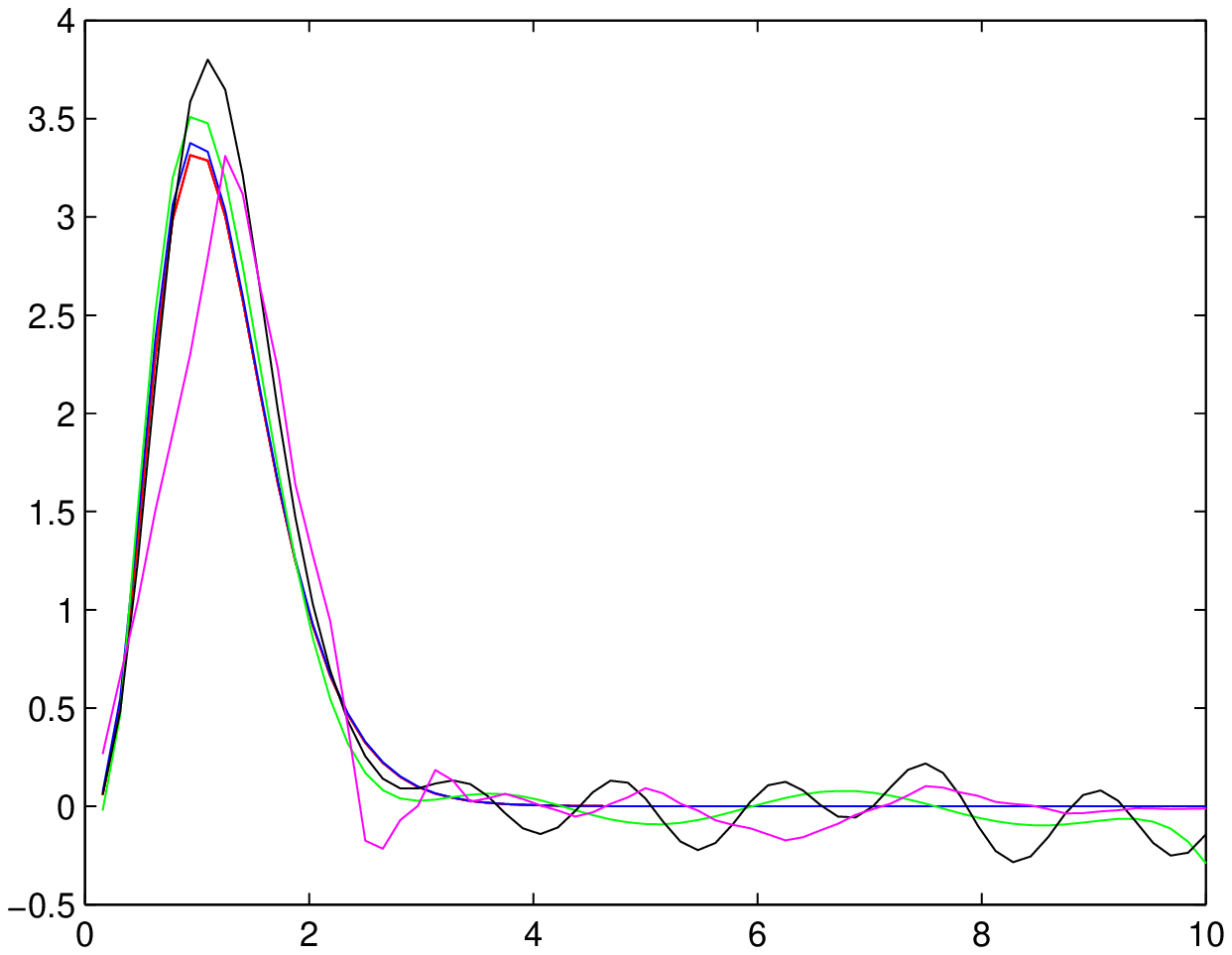} \]
\caption{ Four  estimators of  function $f$ (red) with $n = 64$: the Lasso estimator $\hbf_{Las,cv}$  with Lasso parameter $\hat{\af}$ 
derived by cross validation (blue), the SVD estimator $\hbf_{SVD}$  (black), the Laguerre function estimator 
$\hbf_{Lag}$   (green) and the wavelet-vaguelette estimator $\hbf_{wav}$ (magenta).   
Top row: $\sigma = 0.5$. Bottom row: $\sigma = 1$.
\label{fig:est64}}
\end{figure}
We constructed   Lasso  estimator  $\hbf_{Las,opt} = \bW \hbte (\af_{\tk})$
with the optimal choice of the penalty parameter.  
In practice, $\bof$ is unavailable and parameter $\al$ is chosen by  cross validation. In particular, 
we estimated $q$ by a projection estimator $\widehat{\bq}$ using Laguerre functions basis 
and  the vector of observations $y$ 
and chose $\alpha = \hat{\al} = \alpha_{\hk}$. Here
\bes
\hk = \arg\min_k     \| \bW \hbte(\af_k) - \hbq \|^2_2 + 2 \sig^2 n^{-1} \hat{p}_k
\ees
where $\hat{p}_k$ is the number of nonzero components of $\hbte(\af_k)$ which, in our case,
coincided with the dimension of the linear space $\bW \hbte(\af_k)$.  
At last, we set $\hbf_{Las,cv} = \hbf(\af_{\hk})$.

We compared the Lasso estimators $\hbf_{Las,opt}$ and $\hbf_{Las,cv}$ with the estimators 
$\hbf_{SVD}$, $\hbf_{wav}$ and $\hbf_{Lag}$  where 
$\hbf_{SVD}$ is recovered by the singular value decomposition (SVD), 
 $\hbf_{wav}$ is the wavelet-vaguelette estimator obtained by using  Daubechies wavelet of degree 6
 and $\hbf_{Lag}$ is constructed by expanding the unknown function $f$ over the system of
Laguerre functions  \fr{lag_fun} with $a=1/2$.
The Laguerre functions  dictionary has been proven to be extremely efficient 
for  Laplace deconvolution (see, e.g., \cite{CPR} and  \cite{weeks}). We used $K_s$  eigenbasis functions
for SVD, $K_L$ Laguerre functions for the Laguerre function solution and 
hard thresholding with threshold $\Lam$ for the wavelet-vaguelette estimator.
In order to simplify our numerical work, for all three estimators, 
the SVD, the wavelet-vaguelette and the Laguerre functions based estimator,  
we used the ``ideal'' parameter choices, selecting parameters $K_s$, $K_L$ and $\Lam$ 
 by minimizing the difference between the respective estimators and 
the true function $\bof$ which is unavailable in a real life setting.
Therefore, precision of the three competitive estimators  is somewhat higher 
than  it would be in a real life situation  where parameters of the methods have to be estimated from data.

\begin{figure} [ht]
\[\includegraphics[height=4.0cm]{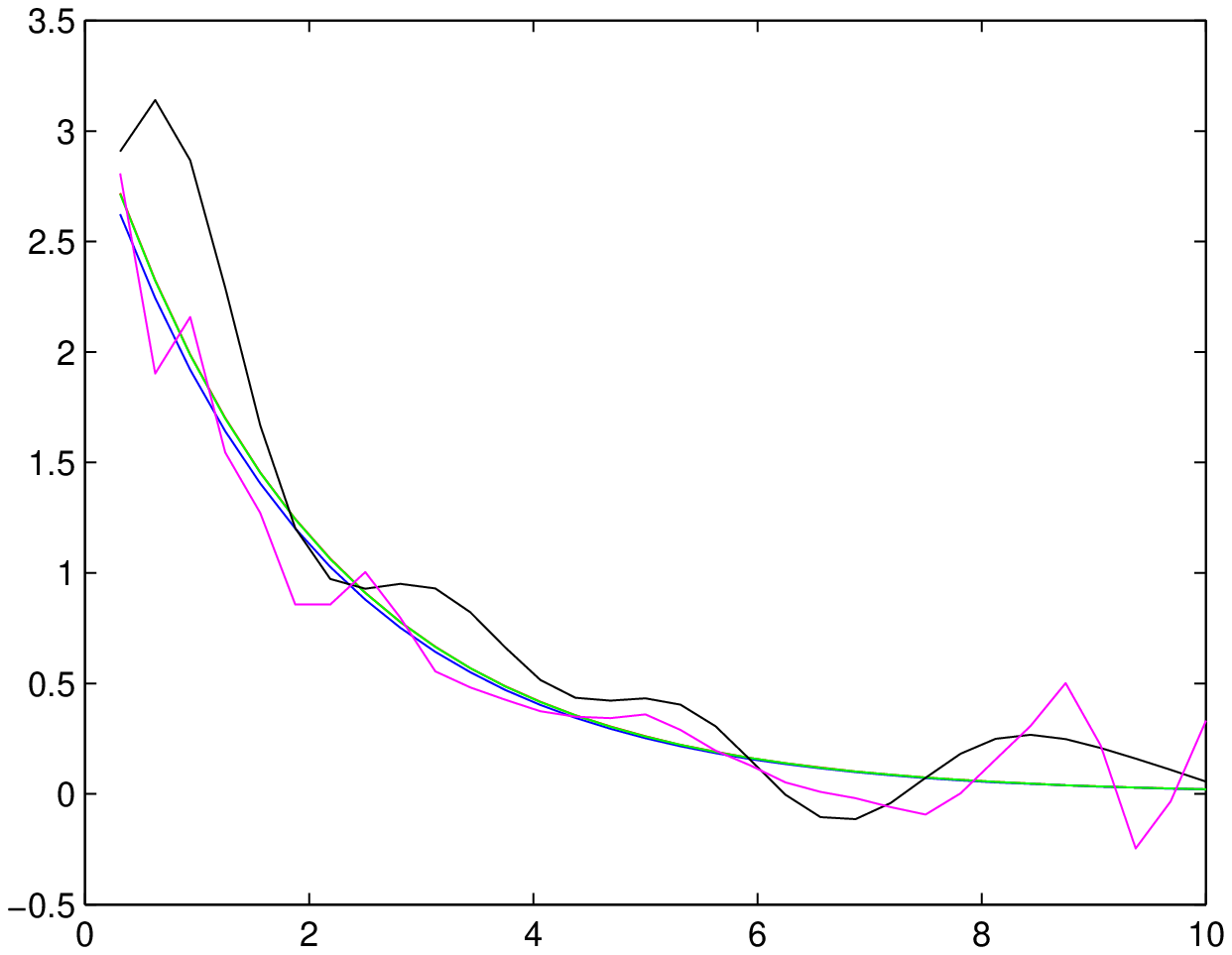} \hspace{2mm}  \includegraphics[height=4.0cm]{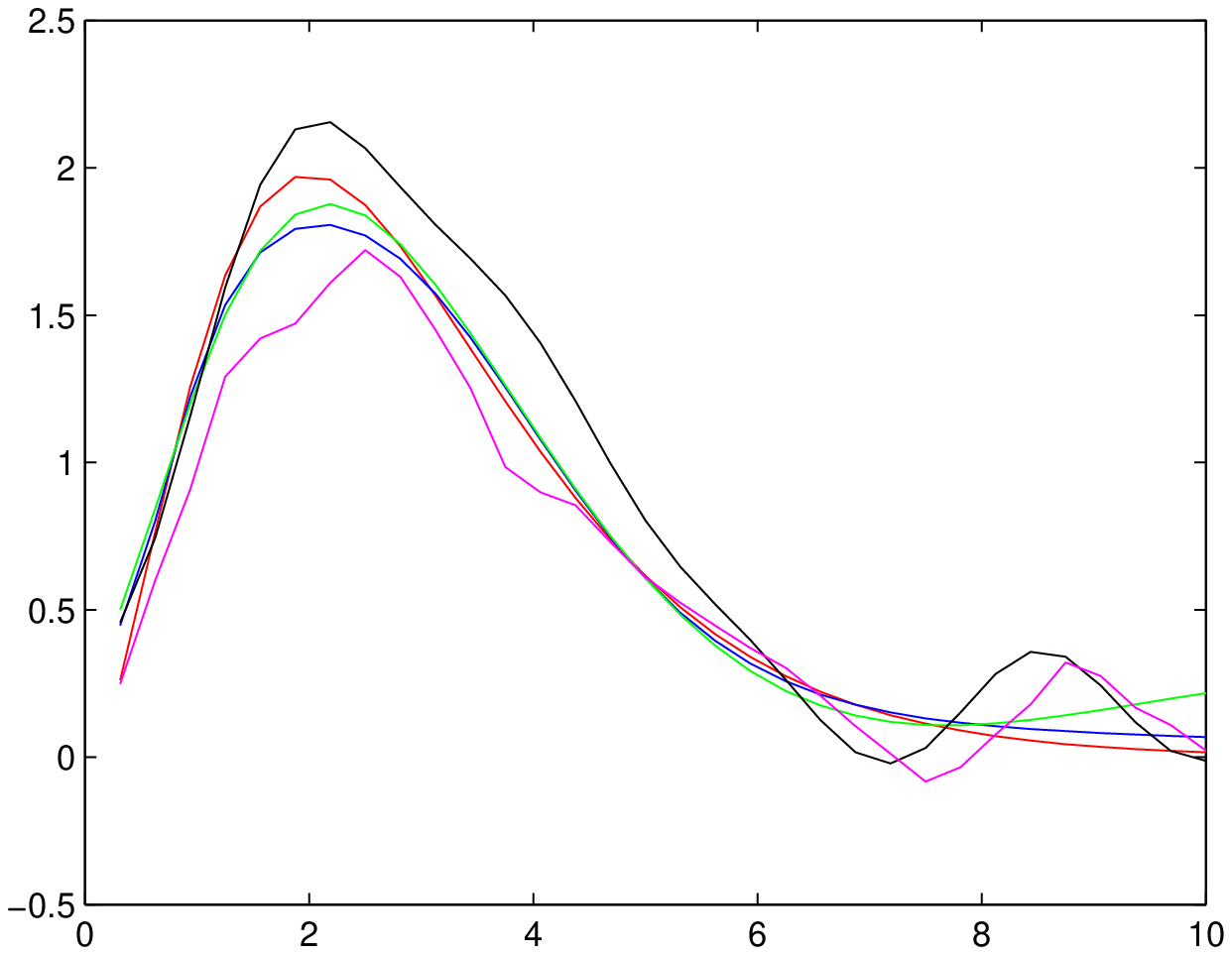}
 \hspace{2mm} \includegraphics[height=4.0cm]{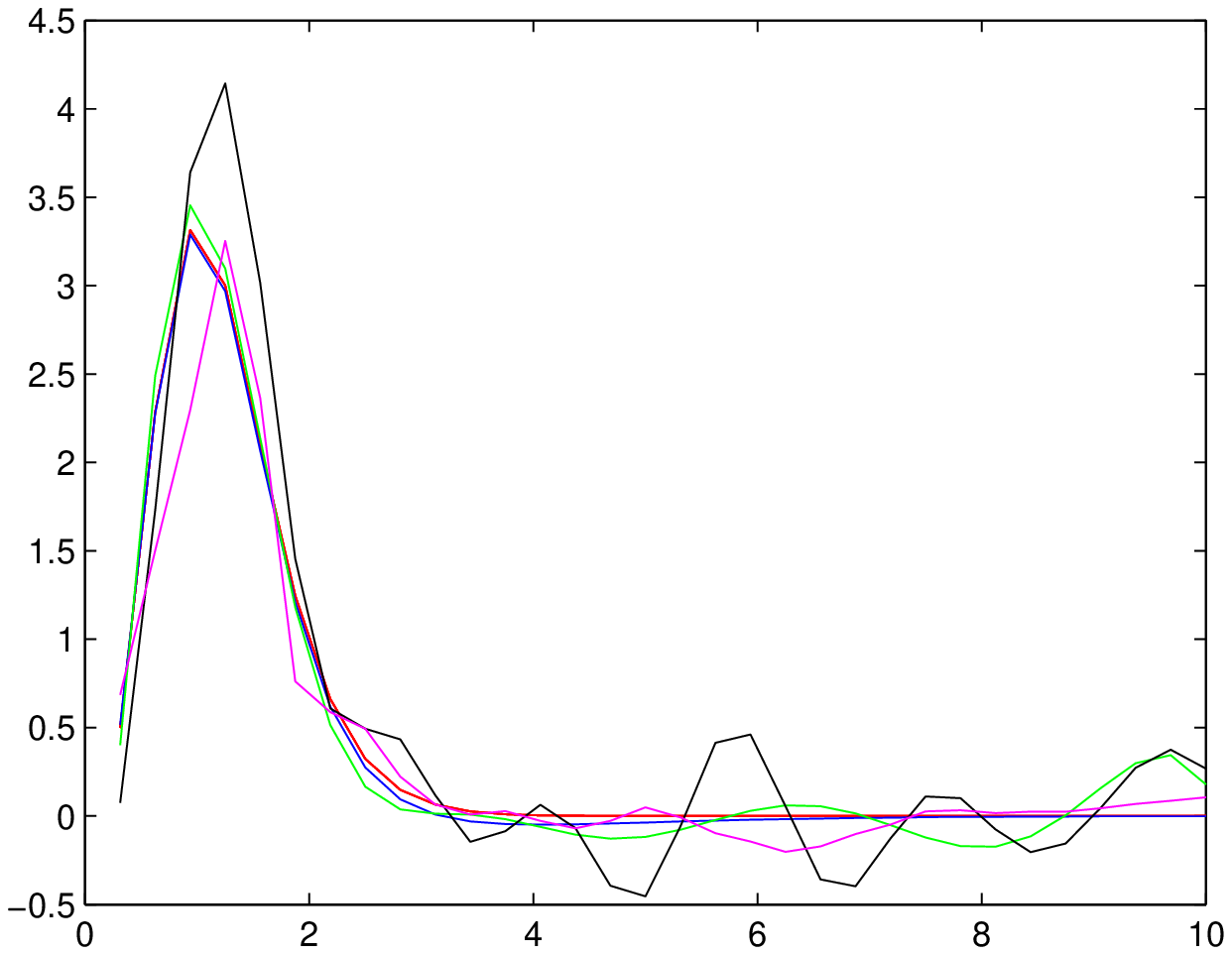} \]
\[\includegraphics[height=4.0cm]{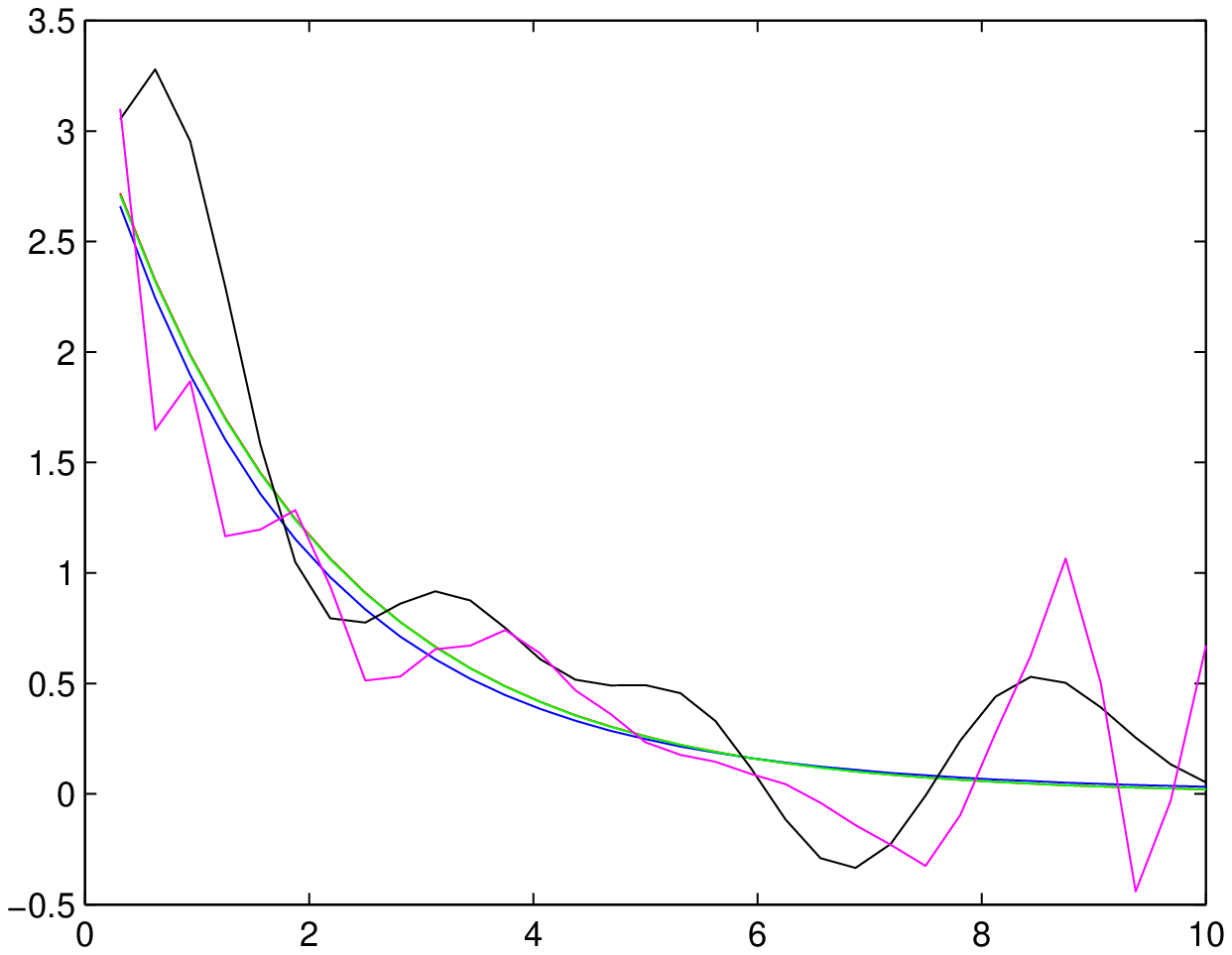} \hspace{2mm}  \includegraphics[height=4.0cm]{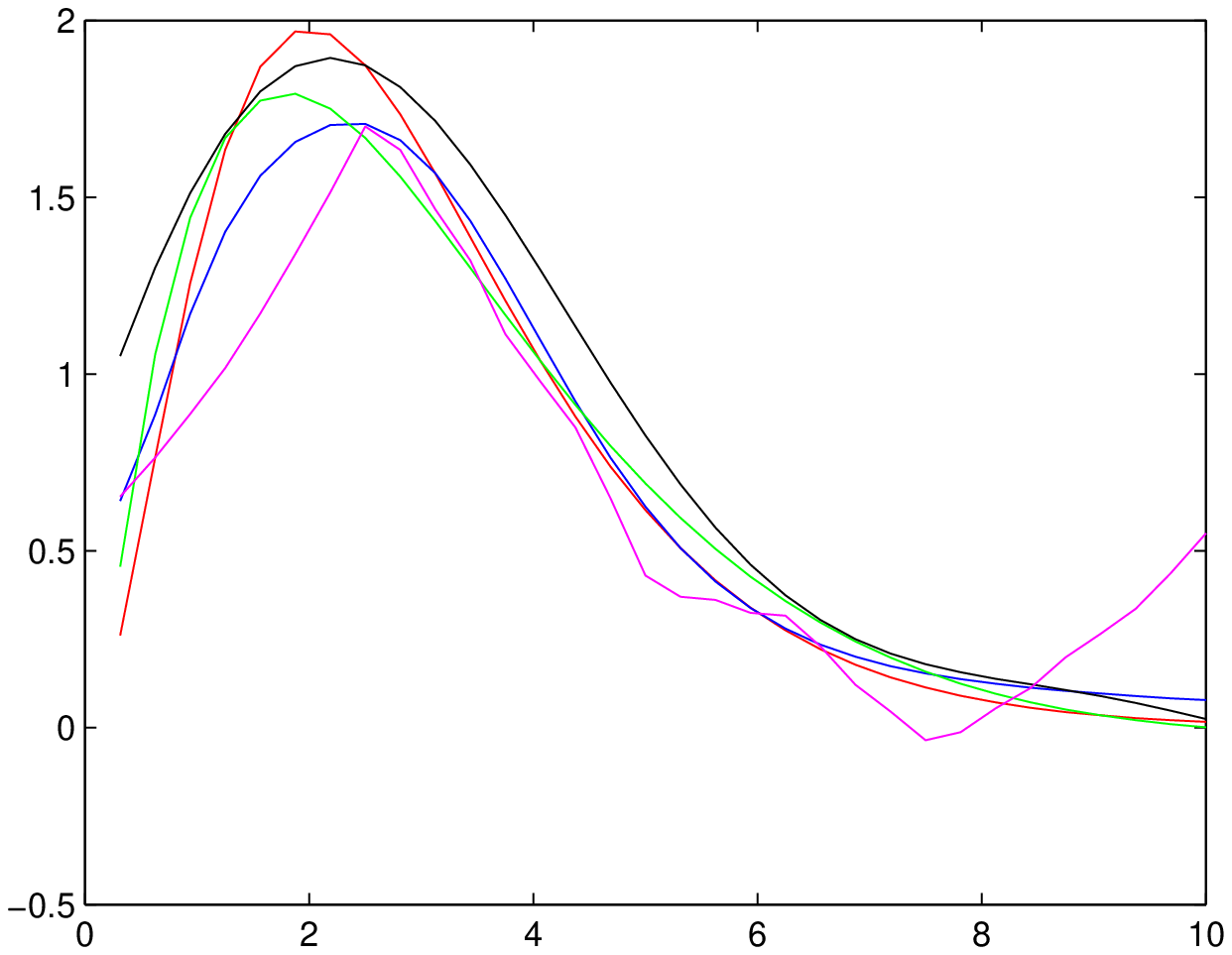}
 \hspace{2mm} \includegraphics[height=4.0cm]{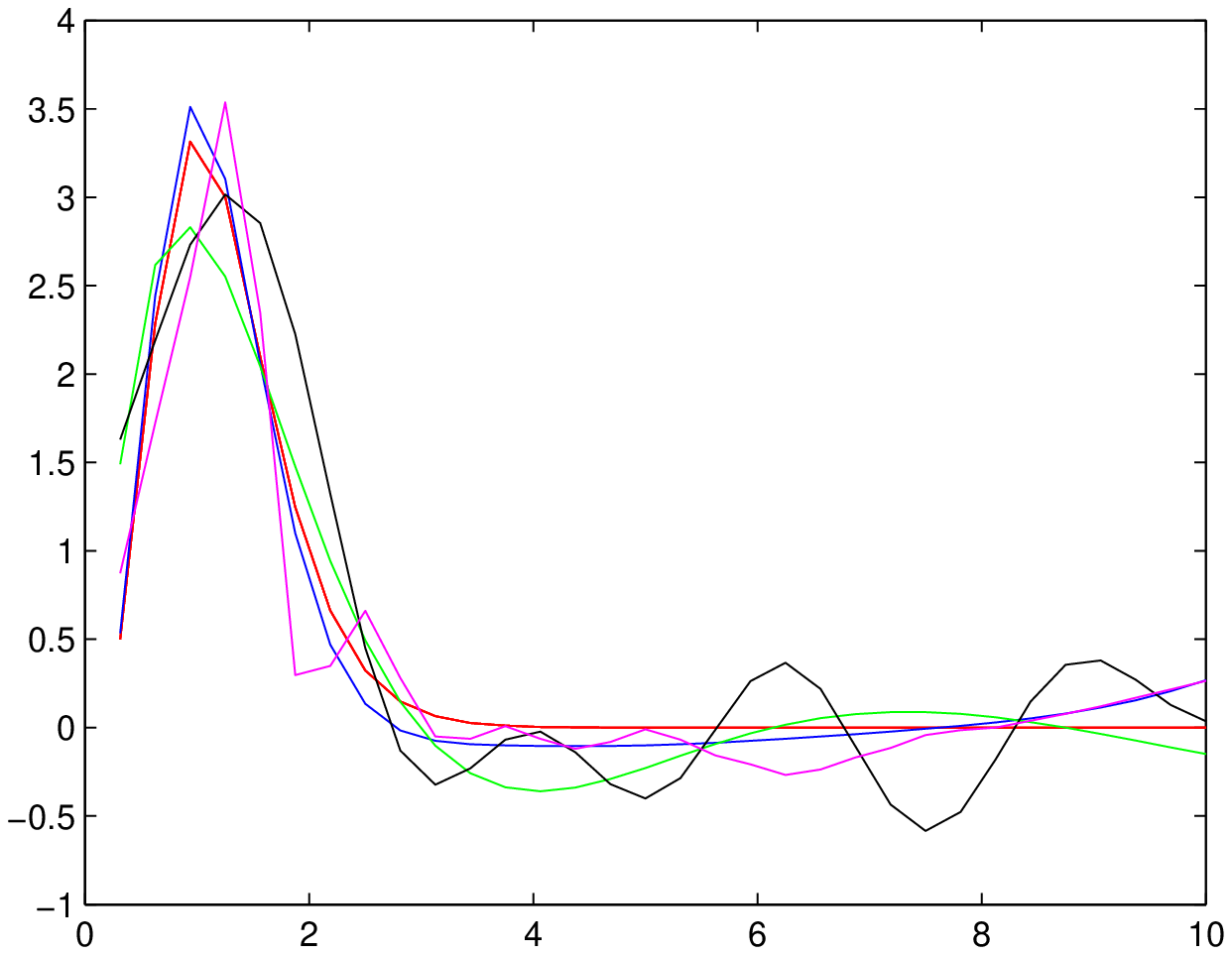} \]
\caption{ Four  estimators of  function $f$ (red) with $n = 32$: the Lasso estimator $\hbf_{Las,cv}$  with Lasso parameter $\hat{\af}$ 
derived by cross validation (blue), the SVD estimator $\hbf_{SVD}$  (black), the Laguerre function estimator 
$\hbf_{Lag}$   (green) and the wavelet-vaguelette estimator $\hbf_{wav}$ (magenta).   
Top row: $\sigma = 0.5$. Bottom row: $\sigma = 1$.
\label{fig:est32}}
\end{figure}

We carried out numerical experiments with two sample sizes, $n=32$ and $n=64$,  three noise levels, 
$\sigma = 0.25$ (low noise level),  $\sigma = 0.5$ (medium noise level) and 
$\sigma = 1$ (high noise level), and several test functions. In particular, we chose three test functions,
 $f_1(x) =  C \exp(-x/2)$, $f_2(x) =  C  x^2*\exp(-x)$  and  $f_3(x) = C  x^4*\exp(-4*x)$,
 where all functions were scaled to have the unit $L^2$ norms. The first test function is easy to estimate and it benefits $\hbf_{Lag}$
since it coincides with the first function of the Laguerre basis. The second function is moderately hard and the last function is 
the most difficult to estimate.

\begin{figure} [ht]
\[\includegraphics[height=4.0cm]{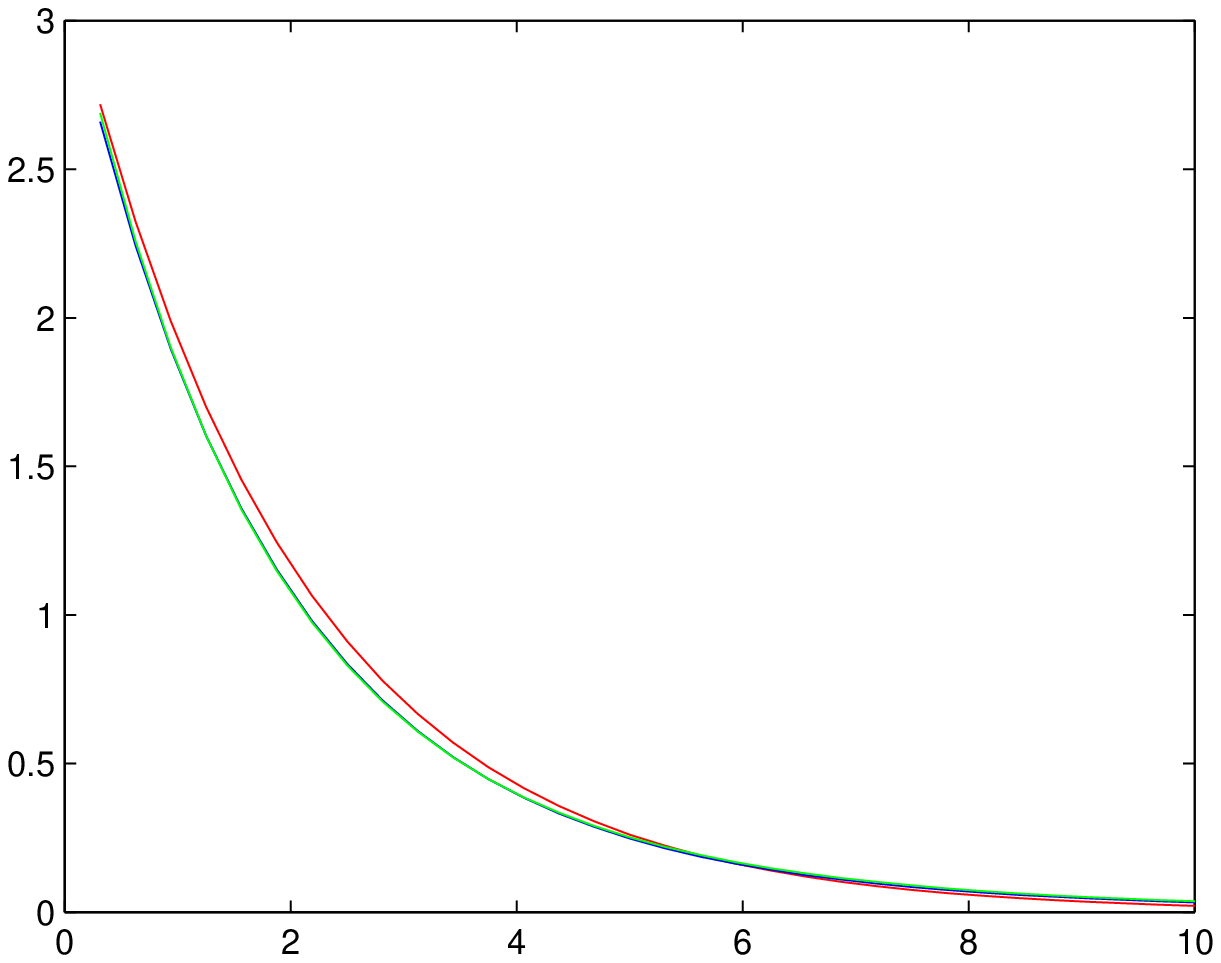} \hspace{2mm}  \includegraphics[height=4.0cm]{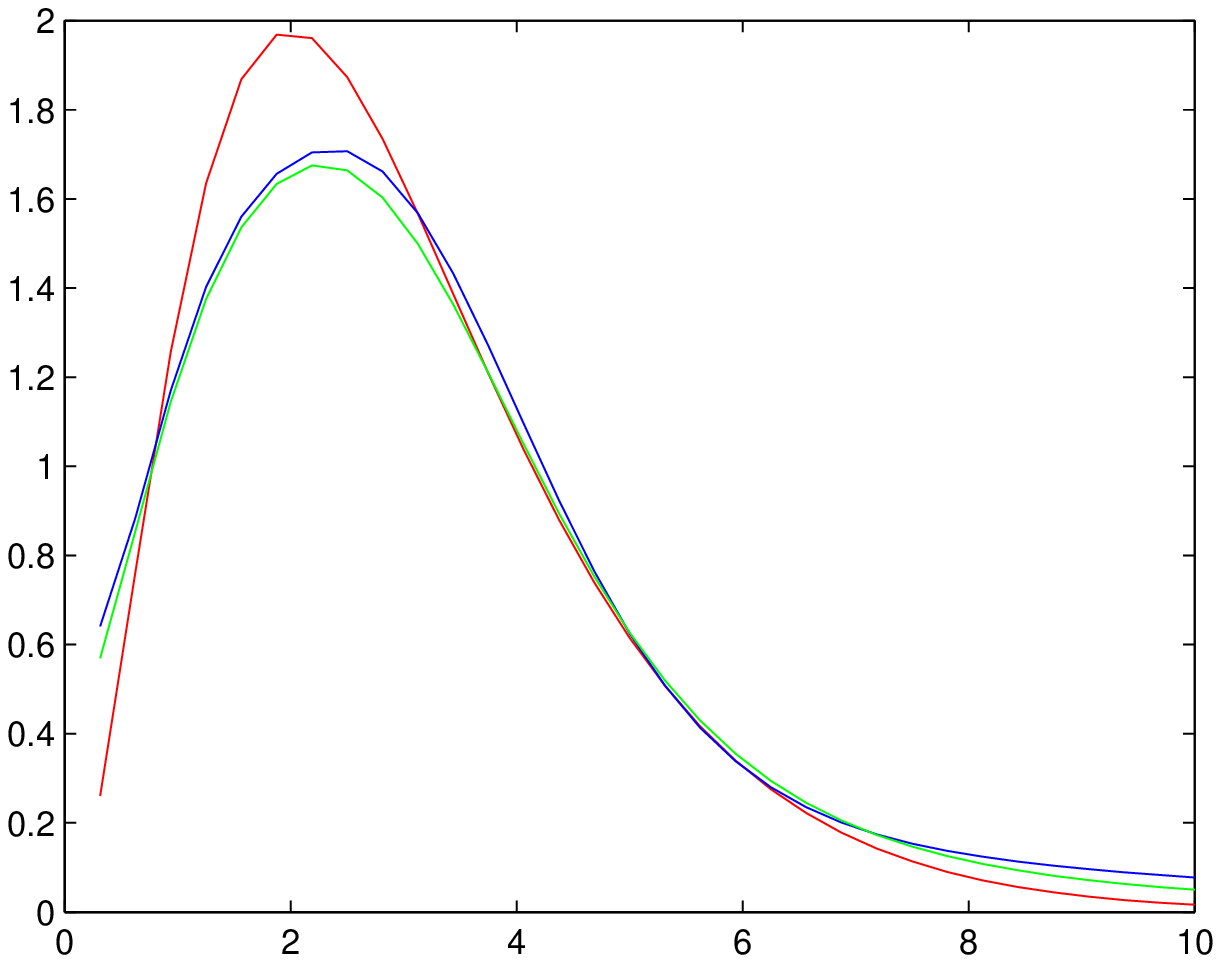}
 \hspace{2mm} \includegraphics[height=4.0cm]{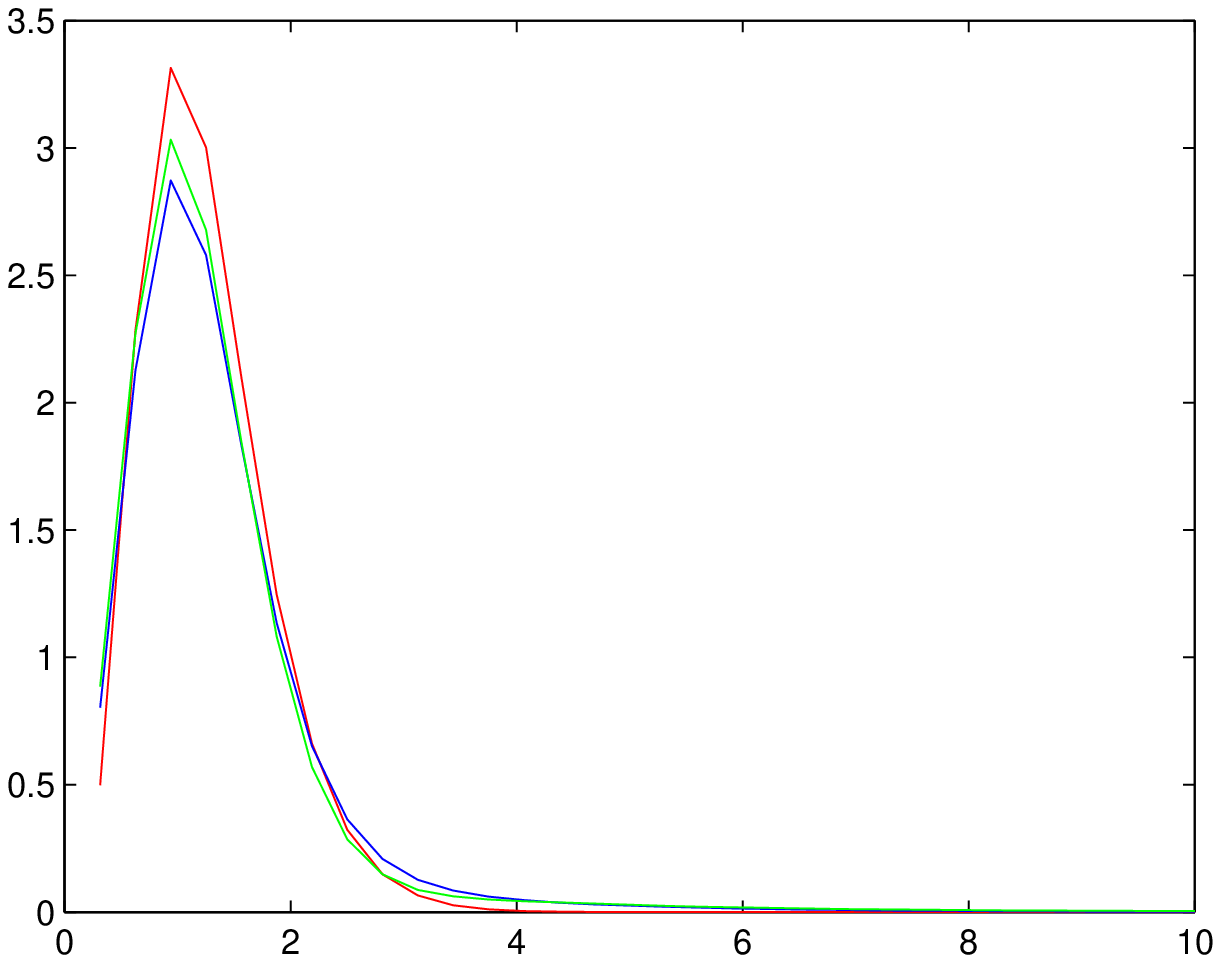} \]
\caption{ Lasso  estimators of  function $f$ (red) with $n = 32$: the Lasso estimator $\hbf_{Las,cv}$ (blue)  with Lasso parameter $\af = \hat{\af}$ 
derived by cross validation and the Lasso estimator $\hbf_{Las,opt}$ (green) with the optimal choice of Lasso parameter.   
\label{fig:cvn32}}
\end{figure}

Figure  \ref{fig:rhs} displays the true functions $q$ and vector $\by$ with $n=32$ and  $\sigma = 0.5$   and $\sigma = 1$, respectively,
for these three cases. Figures~\ref{fig:est64}~and~\ref{fig:est32}  show  the true function $f$ (red) and 
its four  estimators for, respectively, $n=64$ and $n =32$: the Lasso estimator  $\hbf_{Las,cv}$ with the Lasso parameter  
derived by cross validation  (blue), the SVD estimator (black), the Laguerre function based estimator (green) and 
the wavelet-vaguelette estimator (magenta).   Finally, Figure~\ref{fig:cvn32} exhibits the Lasso estimators $\hbf_{Las,opt}$ 
with the optimal choice of parameter $\af$, and $\hbf_{Las,cv}$,
with   parameter $\af$ obtained using cross-validation, when $n=32$ and $\sig =1$ for the three choices of test functions. 
Figure~\ref{fig:cvn32} as well as Table~\ref{table1} below   show that  the  estimators  $\hbf_{Las,opt}$
and $\hbf_{Las,cv}$ (and, consequently, their errors) are very  close to each other.

Table \ref{table1} below compares the accuracy  of the Lasso estimators with the three competitive estimators: 
the SVD estimator, the wavelet-vaguelette
estimator  and Laguerre functions expansion based estimator described above.
Precision of an estimator $\hbf$ is measured by the estimated $L^2$-norm $R(\hbf) =    n^{-1/2}\,  \|\hbf - \bof\|_2$
 of the difference between the estimator $\hbf$ and the true vector $\bof$
averaged over 50 simulation runs  (with the  standard deviations listed  in parentheses).
Columns  1 and 2  present, respectively,  the average MSEs of  the Lasso estimators $\hbf_{Las,opt}$ 
with the optimal choice of parameter $\af$, and $\hbf_{Las,cv}$,
with   parameter $\af$ obtained using cross-validation. Columns 3--5 display the  average MSEs of 
the SVD estimator $\hbf_{SVD}$, the wavelet-vaguelette estimator $\hbf_{wav}$ and the 
Laguerre functions based estimator $\hbf_{Lag}$.


  \begin{table} 
\begin{center}
\begin{tabular}{|l| c |c |c| c| c| }
  \multicolumn{6}{  c  }{{\sc  The accuracies of  the Lasso estimators, the SVD estimators,}}\\
 \multicolumn{6}{ c }{{\sc  the wavelet-vaguelette estimator and the Laguerre functions }}\\
 \multicolumn{6}{ c }{{\sc   based estimators averaged over 50 simulation runs  }}     \\
\hline
Method  & $ \hbf_{Las,opt}$ & $ \hbf_{Las,cv}$ & $\hbf_{SVD}$  & $\hbf_{wav}$ & $\hbf_{Lag}$ \\
\hline
 \hline
$f(x) = \exp(-x/2)$           & 0.019795      &  0.021403   &  0.094387    &  0.106268    &   0.003351   \\
  $\sigma = 0.25$, $n = 64$   & (0.009531)    & (0.011563) & (0.004851)   & (0.014351)   &   (0.002870) \\
\hline
 $f(x) = \exp(-x/2)$         &  0.032464   &   0.038243 &  0.117718   &  0.191543   &   0.006703   \\
  $\sigma = 0.5$, $n = 64$   & (0.017158)  & (0.020279) & (0.009292)  & (0.054438)  &  (0.005740)  \\
\hline
$f(x) = \exp(-x/2)$      &  0.058041  &   0.067571 &  0.153829    & 0.345168    & 0.013406     \\
  $\sigma = 1$, $n = 64$ & (0.031136) & (0.034896) & (0.017815)   & (0.048606)  & (0.011481)  \\
 \hline
 \hline
$f(x) = \exp(-x/2)$           & 0.045767      &  0.048617   &  0.159040    &  0.118117    &   0.007028   \\
  $\sigma = 0.25$, $n = 32$   & (0.016654)    & (0.016153) & (0.007930)   & (0.020318)   &   (0.005496) \\
\hline
 $f(x) = \exp(-x/2)$         &  0.062453   &   0.066710 &  0.185882   &  0.187190   &   0.014057   \\
  $\sigma = 0.5$, $n = 32$   & (0.028558)  & (0.028691) & (0.015800)  & (0.038958)  &  (0.010991)  \\
\hline
$f(x) = \exp(-x/2)$      &  0.100206  &   0.113646 &  0.233622    & 0.363585    & 0.028113     \\
  $\sigma = 1$, $n = 32$ & (0.049762) & (0.050458) & (0.027932)   & (0.077715)  & (0.021983)  \\
 \hline
 \hline
  \multicolumn{6}{ l }{  }\\
\hline
 \hline
 $f(x) = x^2 \exp(-x)$        &  0.015049  & 0.016560   & 0.062621    & 0.090289   &  0.018849    \\
  $\sigma = 0.25$,  $n = 64$  & (0.005400) & (0.005958) & (0.003635)  & (0.005864) &  (0.004934)  \\
\hline
 $f(x) = x^2 \exp(-x)$      &  0.027391  & 0.034609 &  0.076469    &  0.100841   &   0.034744 \\
  $\sigma = 0.5$, $n = 64$  & (0.010565) & (0.016077) & (0.007881)   & (0.009059)  &  (0.010220) \\
\hline
$f(x) = x^2 \exp(-x)$     &  0.051201  &   0.061012 &  0.103803   &  0.143001  & 0.064760  \\
  $\sigma = 1$, $n = 64$  & (0.019715) & (0.022200) &(0.012547)  & (0.027588) & (0.019903) \\
 \hline
 \hline
 $f(x) = x^2 \exp(-x)$        &  0.028626  & 0.031445   & 0.117126    & 0.162770   &  0.035111    \\
  $\sigma = 0.25$,  $n = 32$  & (0.012081) & (0.013028) & (0.005548)  & (0.009057) &  (0.011335)  \\
\hline
 $f(x) = x^2 \exp(-x)$      &  0.046965  & 0.053782 &  0.136472    &  0.181601   &   0.062855 \\
  $\sigma = 0.5$, $n = 32$  & (0.020748) & (0.025215) & (0.012993)   & (0.019603)  &  (0.017499) \\
\hline
$f(x) = x^2 \exp(-x)$     &  0.083337  &   0.095734 &  0.174761   &  0.232028  & 0.104431  \\
  $\sigma = 1$, $n = 32$  & (0.036605) & (0.038061) &(0.024907)  & (0.040098) & (0.022430) \\
 \hline
 \hline
 \multicolumn{6}{ l }{  }\\
\hline
 \hline
 $f(x) = x^4 \exp(-4x)$      &  0.025236  &   0.026016 & 0.095819    & 0.150802   &  0.045406   \\
  $\sigma = 0.25$, $n = 64$  & (0.011729) & (0.011917) & (0.004939)  & (0.015667) &  (0.010733) \\
\hline
$f(x) = x^4 \exp(-4x)$      &  0.040865  &  0.046998 &  0.120739   &  0.211242  &  0.074079   \\
  $\sigma = 0.5$, $n = 64$  & (0.021255) & (0.025032)& (0.012737)  & (0.035823) &  (0.018005) \\
\hline
$f(x) = x^4 \exp(-4x)$    & 0.070826   &  0.085365 &  0.176209  &  0.281932  &  0.126000 \\
  $\sigma = 1$, $n = 64$  & (0.038718) & (0.040451)& (0.024383) & (0.025090) & (0.031723) \\
 \hline
 \hline
 $f(x) = x^4 \exp(-4x)$      &  0.051191  &   0.052743 & 0.180319    & 0.266600   &  0.071385   \\
  $\sigma = 0.25$, $n = 32$  & (0.024067) & (0.023798) & (0.008549)  & (0.021389) &  (0.019737) \\
\hline
$f(x) = x^4 \exp(-4x)$      &  0.071112  &  0.077483 &  0.217531   &  0.311368  &  0.119478   \\
  $\sigma = 0.5$, $n = 32$  & (0.041662) & (0.041242)& (0.023746)  & (0.024043) &  (0.030617) \\
\hline
$f(x) = x^4 \exp(-4x)$    & 0.110801   &  0.128099 &  0.300436  &  0.360937  &  0.204710 \\
  $\sigma = 1$, $n = 32$  & (0.067313) & (0.064436)& (0.049631) & (0.036405) & (0.047348) \\
 \hline
 \hline
\end{tabular}
\end{center}
\caption{ The average values (evaluated over 50 simulation runs) 
of   the errors of the Lasso estimators $\hbf_{Las,opt}$ and $\hbf_{Las,cv}$
the SVD estimator $\hbf_{SVD}$, the wavelet-vaguelette estimator $\hbf_{wav}$  
 and the Laguerre function estimator $\hbf_{Lag}$. 
Standard deviations of the errors are listed in the parentheses. } \label{table1}
\end{table}


Results in Table \ref{table1} confirm that procedure developed in the paper  has good 
computational properties. Indeed, in our simulations, even with the penalty parameter 
obtained via cross validation, Lasso yields better precision 
than both the SVD and the wavelet-vaguelette estimators $\hbf_{SVD}$ and $\hbf_{wav}$
with hand-chosen parameter values.
For the  first test functions,   $\hbf_{Lag}$ is more accurate since the Laguerre basis
contains the function of interest and, due to   orthonormality of the basis, one does not have to pay a price
for selecting the correct dictionary functions. However, for the first test function, 
both  $\hbf_{Lasso}$ and  $\hbf_{Lag}$ produce almost a perfect reconstruction 
as the left panels of Figures~\ref{fig:est64}~and~\ref{fig:est32}  demonstrate.  For the  second and the third test functions,
Lasso exhibits better  precision than its competitors in spite of the fact that we used a fully adaptive Lasso
estimator (the choice of Lasso parameter was data-driven) while, for all three other methods, 
parameters were determined on the basis of the true function $f$ which is not known in practice.

One can easily see that the SVD estimators have relatively high errors. The latter 
can be  explained by the fact that the eigenfunctions of the Laplace convolution operator 
exhibit oscillatory behavior (which can be see on Figures \ref{fig:est64} and \ref{fig:est32}),
so, they require a large number of elements  for representation of   $f$.
The large errors of the wavelet-vaguelette estimator are partially due to using
 periodic wavelets defined on a finite interval   while 
estimating non-periodic functions. Consequently, the wavelet-vaguelette estimator exhibits
strong boundary effects since we did not carry out the boundary correction.


\section{Discussion} 
\label{sec:discussion}
\setcounter{equation}{0}

In the present paper, we consider  application of Lasso to a general linear inverse problem.
The approach is based on inverting of each of the dictionary functions  and matching the resulting expansion   
to the true function $f$. We investigate  the white noise formulation of the problem and further  
extend  the theory to the case of discrete observations with Gaussian or sub-Gaussian noise. 
In addition,   we explain how this methodology can be used when the inverse images of the dictionary functions   
are replaced by their approximate versions. We also show how the technique suggested in the paper can be extended to
the problem of estimation of a mixing density in a continuous mixture.

Using an example of the Laplace convolution equation, we study performance of the Lasso-based estimators via simulations  
and compare their precisions with the SVD estimators, the wavelet-vaguelette estimators and the estimators 
based on the expansion of the unknown function via the Laguerre functions basis. We show that as long as the 
function of interest $f$ has a compact representation in the overcomplete dictionary, the Lasso estimator yields 
satisfactory reconstruction. 
Indeed, in our simulation study, it  demonstrates comparable or better precision 
than its competitors.

Although in the paper we assume that the linear operator $Q$ is completely known, the theory can be extended to the case when operator $Q$
is measured with error or is estimated from the data. The advantage of the approach of the paper is that
it naturally partitions the problem of solution of a linear inverse problem with a noisy operator and 
a right hand side measured with error into two easier problems: solution of an inverse linear problem with the 
noisy operator and   completely known right hand side, and estimation of the linear functional of the 
right hand side on the basis of its noisy version. 
However, solution of   general linear ill-posed problems with noisy operators 
 lie   outside the scope of the present paper and will be treated in future.

\section*{Acknowledgments}

Marianna Pensky   was  partially supported by National Science Foundation
(NSF), grants   DMS-1106564 and DMS-1407475. The author would also like to thank SAMSI for providing support 
which allowed the author's participation in the 2013-14 LDHD program which was instrumental for 
writing this paper.



\section{Proofs }  
\label{sec:proofs}
\setcounter{equation}{0}

Validity  of Theorems \ref{th:slow_Lasso}--\ref{th:mix_den_Lasso} rely on the following Lemma, the proof of which follows the lines 
of reasoning in \cite{arnak}. However, since we are interested in weighted Lasso and 
allow for non-centered errors, for completeness, we provide the proof of the Lemma below.
\\


\begin{lemma} \label{lem:weighted_Lasso}
 Let $f$ be the true function and $\fte$ be its projection onto the linear span of the dictionary $\calL_{\calP}$.
Consider solution of the weighted Lasso problem \fr{las_sol1} with  $\bPhi = \bW^T \bW$, $\bobeta = \bPhi \bte$
and $\hbobeta = \bW^T \bgamma$. Let  
\be \label{bobeta_cond}
\hbobeta = \bobeta + \sqrt{\eps}\bUp \boeta + \bh, \quad \boeta, \bh \in \RR^p,
\ee
where $\EE \boeta =0$ and   components $\eta_j$ of $\boeta$ are sub-Gaussian  random variables satisfying, 
for some $K>0$ and any $t$,
\be \label{large_devK}
\PP \lkr |\eta_i| > t \rkr \leq 2\,\exp(-t^2/K^2).
\ee 
Choose $\tau >0$ and denote
\be \label{ChCalp}
C_h = \max_{1 \leq j \leq p} \lkv \frac{|h_j|}{ \nu_j\, \sqrt{\eps\, \log p}}\rkv, \quad \quad
C_{\al}  = K \sqrt{\tau +1} + C_h.
\ee
If $\al_0 = C_{\af} \sqrt{\eps \log p}$, then for any $\tau >0$  
and any $\af \geq \alfo$, with probability at least $1 - 2 p^{-\tau}$,  one has
\be \label{slowlaslem}
\| f_{\hbte} - f \|_2^2 \leq \inf_{\bt } \lkv  \| f_{\bt } - f \|_2^2 + 4 \al \| \bUp \bt \|_1 \rkv.
\ee
Moreover, if  Assumption  {\bf  A } holds and   $\af = \alfo (\mu +1)/(\mu -1)$, then,  
for any $\tau >0$   with probability at least $1 - 2 p^{-\tau}$,  one has
\be  \label{fasrlaslem}
\| f_{\hbte} - f \|_2^2   \leq   \inf_{\bt, J \subseteq \calP} \lkv  \| f_{\bt } - f \|_2^2 + 4 \al  \| (\bUp \bt)_{\Jc} \|_1
+   \frac{4 C_{\af}^2  \mu^2}{(\mu-1)^2 \kappa^2 (\mu, J)}  \eps \log p \  \sum_{j \in J} \nu_j^2 \rkv.  
\ee 
\end{lemma}

\noindent
{\bf Proof of Lemma \ref{lem:weighted_Lasso}  }. 
Following Dalalyan {\it et al.} (2014), by K-K-T condition, we derive for any $\bt \in \RR^p$
\beqns
\hbte^T (\hbobeta - \bPhi \hbte) & = & \af \| \bUp \hbte \|_1,   \quad
\bt^T (\hbobeta - \bPhi \hbte)   \leq   \af \| \bUp \bt \|_1, 
\eeqns
so that, subtracting the first line from the second,  we obtain
\be \label{main_ineq}
(\hbte - \bt)^T (\bPhi \hbte - \hbobeta) \leq \af \lkr \| \bUp \bt \|_1 - \| \bUp \hbte \|_1 \rkr.  
\ee 
Since $\bPhi \bte = \bobeta$, \fr{main_ineq} yields
\bes 
(\hbte - \bt)^T  \bPhi (\hbte - \bte) \leq \sqrt{\eps}  (\hbte - \bt)^T \bUp \boeta + (\hbte - \bt)^T \bh + 
\af  \lkr \| \bUp \bt \|_1 - \| \bUp \hbte \|_1 \rkr..
\ees
Since for any $\bu, \bv \in \RR^p$ one has 
$
\bv^T \bPhi \bu   =   \frac{1}{2} \lkv \bv^T \bPhi \bv + \bu^T \bPhi \bu - (\bv - \bu)^T \bPhi (\bv - \bu) \rkv,
$
choosing $\bv = \hbte - \bt$ and $\bu = \hbte - \bte$ and observing that for any $\bt$ (and, in particular, for $\bt = \hbte$),
$\| f_{\bt} - f \|_2^2   =   (\bt - \bte)^T \bPhi (\bt - \bte) + \| f_{\bt} - f \|_2^2$,
for any $\bt  \in \RR^p$, one obtains
\be \label{ineq1} 
\| f_{\hbte} - f \|_2^2 + (\hbte - \bt)^T  \bPhi (\hbte - \bte) \leq \| f_{\bt} - f \|_2^2  + 
\sqrt{\eps}  (\hbte - \bt)^T \bUp \boeta + (\hbte - \bt)^T \bh + 2 \af \lkr \| \bUp \bt \|_1 - \| \bUp \hbte \|_1 \rkr.
\ee 
By setting $t = K \sqrt{(\tau +1) \log p}$ in \fr{large_devK} and using \fr{ChCalp},
observe that,  on the set 
\be \label{eq:large_dev} 
\Om = \lfi \om: \max_{1 \leq j \leq p}  |\eta_j|  \leq K \sqrt{(\tau +1) \log p} \rfi \quad \mbox{with}
\quad \PP(\Om) \geq 1 - 2 p^{- \tau}
\ee 
one has 
\bes 
\left| \sqrt{\eps}  (\hbte - \bt)^T \bUp \boeta + (\hbte - \bt)^T \bh \right| 
\leq \sqrt{\eps \, \log p}\, (K \sqrt{\tau +1} + C_h) \, \| \bUp( \hbte - \bt \|_1 
  = \af_0 \, \| \bUp( \hbte - \bt \|_1. 
\ees
Combining the last inequality with \fr{ineq1},  
obtain that, for any $\af >0$, on the set $\Om$,
\be \label{ineq2} 
\| f_{\hbte} - f \|^2 + (\hbte - \bt)^T  \bPhi (\hbte - \bt) \leq \| f_{\bt} - f \|^2
+ 2 \af \lkr \| \bUp \bt \|_1 - \| \bUp \hbte \|_1 \rkr + 2 \alfo \| \bUp( \hbte - \bt) \|_1.
\ee 
Application of inequality  $ \| \bUp( \hbte - \bt) \|_1 \leq  | \bUp \bt \|_1 + \| \bUp \hbte \|_1$  
combined with $\af \geq \af_0$ completes the proof of inequality \fr{slowlaslem}.
\\

In order to prove inequality \fr{fasrlaslem}, denote $\bd = \hbte - \bt$ and observe that, due to $|t_j| - |\hte_j| \leq |\hte_j - t_j|$ and 
$|\hte_j| \geq |\hte_j - t_j| - |t_j|$, inequality \fr{ineq2} implies that, for any set $J \subseteq \calP$, one obtains
\be \label{ineq3}
\| f_{\hbte} - f \|_2^2 + \bd^T  \bPhi \bd \leq \| f_{\bt} - f \|_2^2 + 4 \af \| (\bUp \bt)_{J^c} \|_1   
+ 2 (\af + \alfo)  \| (\bUp \bd)_{J} \|_1   -  2 (\af - \alfo) \| (\bUp \bd)_{J^c} \|_1 .
\ee
Let $\af =  \alfo \, (\mu +1) /(\mu -1)$, so that $(\af + \alfo)/(\af - \alfo) = \mu$.
Now, we consider two possibilities.
If $ \mu\, \| (\bUp \bd)_J \|  <   \| (\bUp \bd)_{J^c} \| $, then 
$
\| f_{\hbte} - f \|_2^2 + \bd^T  \bPhi \bd \leq \| f_{\bt} - f \|_2^2 + 4 \af \| (\bUp \bt)_{J^c} \|_1   
$
and \fr{fasrlaslem} is valid. Otherwise, $\bd \in \calJ (\mu, J)$ and, due to compatibility condition \fr{comp}
and inequality $2 a b \leq a^2 + b^2$, one derives
\bes
2 (\af + \alfo)\ \| (\bUp \bd)_{J} \|_1 \leq 2 (\af + \alfo)\  \sqrt{\Tr(\bUp_J^2)\,  \bd^T \bPhi \bd} /\kappa (\mu, J)
\leq \bd^T \bPhi \bd + (\af + \alfo)^2\,   \Tr(\bUp_J^2)/\kappa^2 (\mu, J).
\ees
Plugging the latter into \fr{ineq3} and using  $\af = \alfo (\mu +1)/(\mu -1)$, obtain that 
\fr{fasrlaslem} holds   for any $\bt$.
\\


\noindent
{\bf Proof of Theorem \ref{th:slow_Lasso}. }\  Let $\bobeta$ and $\hbobeta$ be the vectors with components $\bobeta_j = \lan f, \ph_j \ran_{\calH_1}$ 
and $\hbobeta_j = \lan   y,   \psij \ran_{\calH_2}$, $j=1, \cdots, p$.
Then, due to \fr{mainrel}, one has  $\hbobeta_j = \bobeta_j + \sqrt{\eps} \nuj \boeta_j$
where $\boeta_j$ are standard normal variables.
Moreover, if   $f$ is the true function and $\fte$ is its projection onto the span of the dictionary $\calL_{\calP}$,
then, $\lan  f- \fte;  \ph_j \ran$ for $j=1, \cdots, p$, and  $\bbe = \bPhi \bte$. 
Therefore, validity of Theorem \ref{th:slow_Lasso} follows from Lemma \ref{lem:weighted_Lasso} with 
$K = \sqrt{2}$, $\bh=0$ and  $C_h =0$  in  \fr{ChCalp}.\\

\medskip


\noindent
{\bf Proof of Theorem \ref{th:fast_Lasso}. }\  Validity of \fr{fasrlas} follows from Lemma \ref{lem:weighted_Lasso} with 
$K = \sqrt{2}$, $\bh =0$  and  $C_h =0$  in  \fr{ChCalp}, so that $C_{\af} = \sqrt{2(\tau +1)}$ and $K_0 =2$ in \fr{fasrlas}. 
In order to prove \fr{corfast}, choose $f_{\bt} = \proj_{\calL _J} f$, so that
$t_j =0$ for $j \in \Jc$. \\

\medskip


 \noindent
{\bf Proof of Theorem \ref{th:obs_Lasso}. }
Note that vector $\hbobeta$ has components $\hbej = \bej + \deljo + \deljt$, $j=1, \cdots, p$,
where
\beqns
\deljo & = & \sumin \xi_i \psij (x_i) \Dxi,\\
\deljt & = &  \sumin q(x_i) \psij (x_i) \Dxi - \int_{\calX} q(x) \psij (x) dx,
\eeqns
are, respectively, the random error component and the bias of $\hbej$.
In order to bound above the random term, apply Proposition 5.10 of \cite{vershynin}
which implies that, for any vector $\ba$ and any $z>0$, one has
\bes
\PP \lkr \left| \sum_{i=1}^n a_i \xi_i \right| > z \rkr \leq e\, \exp \lkr -\frac{z^2}{2 \sig^2 \| \ba \|_2^2} \rkr.
\ees 
Choosing $a_j = \psi_j (x_i) \Dxi$ and $z =  \sig \nu_j t/\sqrt{n}$ and noting that, by assumption \fr{alephcond}, 
one has $\| \ba \|^2_2 \leq n^{-1} \nuj^2 \vart^2$,  obtain 
$$
\PP \lkr  |\deljo| > \frac{ \sig \, \nuj t}{\sqrt{n}} \rkr \leq e\, \exp\lfi - \frac{t^2}{2 \vart^2} \rfi.
$$
Also, it is known that the error of the rectangular approximation of an integral obeys
$|\deljt| \leq \nuj \aleph  T^2/(2n).$
 Apply  Lemma \ref{lem:weighted_Lasso}
with $\eps = \sig^2/n$, $K = \vart \sqrt{2}$, $h_j = \deljt$ and 
$C_h = T^2 \vart \aleph /(2 \sig \sqrt{n \, \log p})$ and observe that 
for $n \geq  \calN$, one has $K \sqrt{\tau +1} \geq C_h$.
Then, for $C_{\af} = 2 \vart \sqrt{2 (\tau +1)}$,   
obtain that, with probability at least $ 1 - e\, p^{- \tau}$, 
inequalities \fr{fasrlas} and \fr{corfast} hold with $K_0 = 8 \vart^2$.
\\

 \medskip


\noindent
{\bf Proof of Theorem \ref{th:mix_den_Lasso}. }\\
To prove the theorem, apply Lemma~\ref{lem:weighted_Lasso} with $\bh =0$
and $\eta_j = \sqrt{n}(\hbej - \bej)/\nuj$ in \fr{bobeta_cond}. 
The main difference between the proof of this theorem and Theorem \ref{th:fast_Lasso}
is that we establish inequality \fr{eq:large_dev} directly instead of relying on assumption 
\fr{large_devK}. For this purpose, we observe that 
\bes 
\eta_j = n^{-1} \sum_{i=1}^n  z_{ij} \quad \mbox{with} \quad 
z_{ij} = \frac{\sqrt{n}}{\nuj} \psij(Y_i) - \frac{\sqrt{n}}{\nuj} \EE \psij(Y_i) 
\ees
with $\EE z_{ij} =0$, 
 $\EE z_{ij}^2   = \sigma^2_z = n \nuj^{-2} \Var [\psij(Y_1)] \leq 1$ and 
$\|z_{ij}\|_\infty = \max  |z_{ij}|   <  2 \sqrt{n} \| \psij \|_\infty/\nuj$.
Applying Bernstein inequality, we obtain
\be \label{bern} 
\PP(|\nuj| > z) \leq 2\ \exp \lfi - \frac{z^2}{2} \, 
\lkr 1 +\frac{2 z \, \| \psij \|_\infty}{3 \sqrt{n} \, \nuj} \rkr^{-1} \rfi.
\ee
Choosing $z = 2\,   \sqrt{(\tau +1) \log p}$ in \fr{bern} and 
noting that $2 z \, \| \psij \|_\infty/(3 \sqrt{n} \, \nuj) \leq 1$ for 
$n \geq \calN_0$, we obtain \fr{eq:large_dev}  with $K = 2$. 
Application of Lemma~\ref{lem:weighted_Lasso} completes the proof. 
\\ 

\medskip
 

\noindent
{\bf Proof of Corollary \ref{cor:wishart}. }
In order to prove validity of the corollary, we just need to verify the expression for $\nuj$ in
\fr{nuj_wishart}. For simplicity, we drop the index $j$.
Observe that since $\bA$ is symmetric and positive definite, there exists a symmetric square root
$\sqrt{\bA} = \bA^{1/2}$ and that expression \fr{psiAgam} can be re-written as 
\bes  
\psi (\bY| \bA, \ga) = \frac{\Gamma_r \lkr \frac{m}{2} \rkr\  2^{\frac{ \ga  r}{2}} |\bA|^{-\frac{r+1}{4}} }
{\Gamma_r \lkr \frac{m-\ga}{2} \rkr\ \sqrt{\Gamma_r \lkr \frac{2\ga - r-1}{2} \rkr}}\ 
\frac{|\bA^{-1/2}\bY \bA^{-1/2} - \bI|^{\frac{m -\ga -r-1}{2}}   }
{|\bA^{-1/2}\bY \bA^{-1/2}|^{\frac{m-r-1}{2}} } \ \II(\bA^{-1/2}\bY \bA^{-1/2} - \bI >0).
\ees
Furthermore, note that matrix $\bA^{-1/2}\bY \bA^{-1/2}$ is symmetric, so that there exists a diagonal matrix $\bD$ 
with components $D_k$, $k=1, \cdots, p$, and 
an orthogonal matrix $\bU$ such that $\bA^{-1/2}\bY \bA^{-1/2} = \bU \bD \bU^T$. Using the fact that 
$|\bU|=1$, obtain that 
\bes 
|\bA^{-1/2}\bY \bA^{-1/2} - \bI|^{\frac{m -\ga -r-1}{2}}\ 
|\bA^{-1/2}\bY \bA^{-1/2}|^{-\frac{m-r-1}{2}} = \prod_{k=1}^p \lkv (D_k-1)^{\frac{m -\ga -r-1}{2}}\  (D_k)^{-\frac{m-r-1}{2}} \rkv.
\ees    
Maximizing the last expression with respect to $D_k >1$, obtain \fr{nuj_wishart}.
\\

 \medskip


\noindent
{\bf Proof of Lemma \ref{lem:coherence}. }
First, we prove that 
\be \label{coh:nondiag}
0 < \ro (l_1, l_2; b_1, b_2) \leq \lkr \frac{b_1}{b_2} \rkr^{2(l_1 - l_2)}\  
\exp \lfi  - \frac{|l_1 - l_2|^2}{2 \max(l_1, l_2)} -  (l_1 + l_2 +1) \, \cosh \lkr \log \sqrt{\frac{b_1}{b_2}} \rkr \rfi. 
\ee
For this purpose, observe that elements of matrix $\bPhi$ are of the form
\bes
\ro (l_1, l_2; b_1, b_2) = \lan \ph_{l_1, b_1}, \ph_{l_2, b_2} \ran =  \sqrt{R_1(l_1, l_2)}\, \sqrt{ R_2(l_1, l_2; b_1, b_2)}
\ees
with 
\bes
R_1(l_1, l_2) = \frac{(l_1 + l_2)!)^2}{(2 l_1)! (2 l_2)!}; \quad
R_2(l_1, l_2; b_1, b_2) = \frac{2^{2 l_1 + 2 l_2 +2} b_1^{2 l_1 +1} b_2^{2 l_2 +1}}
{(b_1 + b_2)^{2 l_1 + 2 l_2 +2}}
\ees
Let $l_1 \neq l_2$. Denote $d = |l_1 -  l_2| = \max(l_1, l_2) -  \min(l_1, l_2)$ and $l =  \min(l_1, l_2)$.
Then, $l_1 + l_2 = 2 l + d$ and, using inequality $\log(1-x) \leq -x$ for $0<x<1$, one derives
\beqn 
R_1(l_1, l_2) & =  & \frac{[(2l+d)!]^2}{(2l + 2d)! (2l)!} = \exp \lfi \sum_{j=1}^d\ \log \lkr 1 - \frac{d}{2l + j + d} \rkr \rfi \nonumber \\
& \leq &  \exp \lfi - \sum_{j=1}^d\  \frac{d}{2l + j + d}   \rfi
\leq \exp \lfi - \frac{d^2}{2(l+d)} \rfi = \exp \lfi - \frac{|l_1 - l_2|^2}{2 \max(l_1, l_2)} \rfi. \label{R1}
\eeqn 
In order to obtain an  upper bound for $R_2(l_1, l_2; b_1, b_2)$, denote
$h = b_1/b_2$. Then,
\beqns 
R_2(l_1, l_2; b_1, b_2) & = & \frac{2^{2 l_1 + 2 l_2 +2}\,  h^{2 l_1 +1} }{(1 +h)^{2 l_1 + 2 l_2 +2}} = 
\lkv \frac{1 + h}{2 \sqrt{h}} \rkv^{-2(l_1 + l_2 +1)} h^{2(l_1 - l_2)}.
\eeqns
Note that 
\bes
\frac{1 + h}{2 \sqrt{h}} = \frac{h^{1/2} + h^{-1/2}}{2} =   \cosh (\log(\sqrt{h})), 
\ees 
so that
\beqn \label{R2} 
R_2(l_1, l_2; b_1, b_2) & \leq & \lkr \frac{b_1}{b_2} \rkr^{2(l_1 - l_2)} 
\exp \lfi  -(l_1 + l_2 +1) \, \cosh \lkr \log \sqrt{\frac{b_1}{b_2}} \rkr \rfi. 
\eeqn 
Combining \fr{R1} and \fr{R2}, obtain  \fr{coh:nondiag}.

Now, it is easy to see that, due  to the fact that $\cosh (x) > \frac{1}{2}\, e^{|x|}$, 
for $\ro$ defined in \fr{nondiag}, one obtains     
\bes  
0 < \ro (l_1, l_2; b_1, b_2) \leq \lkr \frac{b_1}{b_2} \rkr^{2(l_1 - l_2)}\  
\exp \lfi  - \frac{|l_1 - l_2|^2}{2 \max(l_1, l_2)} -    
  \lkr \frac{(l_1 + l_2 +1)}{2}\rkr\,  \lkv \left| \log \lkr \frac{b_1}{b_2} \rkr  \right| - \log 4 \rkv \rfi.
\ees
Therefore,   for any pair of indices $j, k \in \calP$ such that $l_j \leq l_k$, inequality \fr{corineq} holds
provided $b_j \geq b_k$.  
\\




\begin{thebibliography}{10}
 
\bibitem{APR}
Abramovich, F., Pensky, M.,  Rozenholc, Y. (2013)
Laplace deconvolution with noisy observations. 
 {\it Electronic Journal of Statistics}, {\bf 7},  1094-1128


\bibitem{abr}     
Abramovich, F.,    Silverman, B. W.  (1998). 
Wavelet decomposition approaches to statistical inverse problems.
{\it Biometrika\ }, {\bf 85},  115--129.


\bibitem{bickel_ritov_tsybakov} 
Bickel, P.J.,  Ritov, Y., Tsybakov, A. (2009) Simultaneous analysis of Lasso and Dantzig selector. 
{\em Ann. Statist.}, {\textbf 37}, 1705 - 1732.




 \bibitem{bissantz}
 Bissantz, N., Hohage, T., Munk, A., and Ruymgaart, F. (2007)  
 Convergence rates of general regularization methods for statistical 
 inverse problems and applications.
 {\it SIAM J. Numer. Anal.}, {\bf 45}, 2610--2636.


\bibitem{sara}
B{\"u}hlmann, P., van de Geer, S. (2011) 
{\em Statistics for High-Dimensional Data: Methods, Theory and Applications.} Springer.


\bibitem{bunea_tsybakov_2}  
Bunea, F.,  Tsybakov, A.,  Wegkamp, M. (2007) Sparsity oracle inequalities for the Lasso. 
{\em Electron. J.Stat.}, \textbf{1}, 169 - 194.


\bibitem{bunea1}
Bunea, F.,   Tsybakov, A.,  Wegkamp, M.,    Barbu, A. (2010)
Spades and Mixture Models.  {\em Ann. Statist.}, {\textbf 38},   2525 - 2558.




\bibitem{cavalgol2}
Cavalier, L., Golubev, Yu. (2006)
Risk hull method and regularization by projections 
of ill-posed inverse problems.
 {\it Ann. Statist.}, {\bf 34},  1653–-1677.


\bibitem{cavalgol1}
Cavalier, L., Golubev, G.K., Picard, D., Tsybakov, A.B.  (2002)  
Oracle inequalities for inverse problems. {\it Ann. Statist.}, {\bf 30},  843-–874.  

\bibitem{cavreiss}
Cavalier, L., Reiss, M. (2014)
 Sparse model selection under heterogeneous noise:
Exact penalisation and data-driven thresholding.
{\it Electronic Journ. Statist.}, {\bf 8}, 432-455.

 \bibitem{cohen}
 Cohen, A., Hoffmann, M., Reiss, M. (2004)
 Adaptive wavelet Galerkin methods for linear inverse problems. 
 {\it SIAM Journ. Numer. Anal.}, {\bf 42}, 1479--1501.


\bibitem{CPR}
Comte, F.,    Cuenod, C.-A.,    Pensky, M.,   Rozenholc, Y. (2015)
Laplace deconvolution on the basis of time domain data and its
application to Dynamic Contrast Enhanced imaging
\texttt{arxiv:~1405.7107.v2}


\bibitem{comte}
Comte, F.,  Genon-Catalot, V. (2015) 
Adaptive Laguerre density estimation for mixed Poisson models.
\texttt{Preprint Hal MAP5 Preprint 2013-15}.  



\bibitem{arnak}
Dalalyan, A.S., Hebiri, M., Lederer, J. (2014)
On the prediction performance of the Lasso.
\texttt{arxiv:~1402.1700}


 


 \bibitem{donoho}
Donoho, D.L. (1995). 
Nonlinear solution of linear inverse problems by wavelet-vaguelette decomposition.   
{\it Appl. Computat. Harmonic Anal.},  {\bf 2} 101--126.


\bibitem{efrom}
Efromovich, S., Koltchinskii, V.   (2001). On inverse problems with unknown operators.
{\it IEEE Trans. Inform. Theory}, {\bf 47},  2876 - 2894.

\bibitem{gol}
Golubev, Y. (2010)
On universal oracle inequalities related to high-dimensional linear models.
 {\it Ann. Statist.}, {\bf 38},   2751--2780. 


  \bibitem{goutis}
  Goutis, C. (1997)
  Nonparametric Estimation of a Mixing Density via the Kernel Method.
 {\it Journ. Amer.Stat. Assoc.}, {\bf 92},  1445-1450.


\bibitem{grad}
Gradshtein, I.S.,  Ryzhik, I.M. (1980) {\em Tables of
integrals, series, and products.} Academic Press, New York.




\bibitem{gripenberg}
 Gripenberg, G., Londen, S.O.,  Staffans, O.  (1990).
{\it Volterra Integral and Functional Equations.}
Cambridge University Press, Cambridge.
 

\bibitem{gupta}
Gupta, A.K.,  Nagar, D. K. (1999).
{\it Matrix Variate Distributions.} 
CRC Press.


\bibitem{hern}
Herngartner, N.W. (1997)
Adaptive demixing in Poisson mixture models.
{\it Ann. Statist.}, {\bf 25}, 917-928.
 
\bibitem{hoffmann}
Hoffmann, M., Reiss, M. (2008)
Nonlinear estimation for linear inverse problems with error in the operator. 
 {\it Ann. Statist.}, {\bf 36}, 310--336. 

\bibitem{kalifa}
Kalifa, J., Mallat, S. (2003). Thresholding estimators for linear
inverse problems and deconvolutions.  {\it Ann. Statist.}, 
{\bf 31} 58--109.

\bibitem{klopp}
Klopp, O., Pensky, M. (2014)
Sparse high-dimensional varying coefficient model:
non-asymptotic minimax study. 
{\em Ann  Statist.}, in press. 



\bibitem{levin}
Liu, L., Levine, M., and Zhu, Y. (2009) 
A Functional EM Algorithm for Mixing Density Estimation via Nonparametric Maximum Likelihood Maximization.
{\it  Journ. Comput.  Graphical Statist.}, {\bf 18},    481--504.



\bibitem{lounici_pontil_tsybakov}
Lounici, K., Pontil, M., Tsybakov, A., van de Geer, S. (2010)  
Oracle inequalities and optimal inference under group sparsity.
{\em Ann  Statist.}, \textbf{39}, 2164-2204.



\bibitem{meister}
Meister, A. (2009)
{\it Deconvolution Problems  in Nonparametric Statistics.}
{\it Lecture Notes in Statistics}, {\bf 193},  Springer-Verlag,  Berlin.



 

\bibitem{handbook}
 Polyanin, A.D.,  Manzhirov, A.V.  (1998).
{\it Handbook of Integral Equations}, CRC Press, Boca Raton, Florida.

\bibitem{spams}
Mairal, J. (2014)
{\it SPAMS: a SPArse Modeling Software}, MatLab toolbox. 
 {\tt http://spams-devel.gforge.inria.fr}

\bibitem{tibsh}
Tibshirani, R. J.,   Taylor, J. (2012) Degrees of freedom in lasso problems. 
{\it Ann. Statist.}, {\bf 40}, 1198 -1232.
 

\bibitem{tropp}
Tropp, J.A.,    Wright, S. J. (2010)
Computational methods for sparse solution of linear inverse problems.
{\it Proc. IEEE,   special issue, "Applications of sparse representation and compressive sensing"},
{\bf 98},   948-958.



 

\bibitem{vershynin}
{Vershynin, R.}  (2012)
{\em Introduction to the non-asymptotic analysis of random matrices.} 
In {\it  Compressed Sensing, Theory and Applications}, ed. Y. Eldar and G. Kutyniok, Chapter 5.
 {Cambridge University Press}. 


\bibitem{walter}
Walter, G.G. (1981)
Orthogonal series estimators of the prior distribution.
{\em Sankhy\={a}}, {\bf A43}, 228-245.

\bibitem{weeks}
Weeks, W.T. (1966)
Numerical Inversion of Laplace Transforms Using Laguerre Functions.
{\it J. Assoc. Comput. Machinery}, {\bf 13},
 419 - 429.


\bibitem{yuan} 
Yuan, M., Lin, Y. (2006) 
Model selection and estimation in regression with grouped variables. 
{\em J. R. Stat. Soc., Ser. B}, \textbf{68},  49 - 67.




 
\end{thebibliography}
\end{document}